\documentclass[aps,preprint]{revtex4}
\usepackage{graphicx}
\begin{document}

\title{Field theory Lagrangian approach to nuclear structure}

\author{Tapas Sil$^1$, S. K. Patra$^2$, B. K. Sharma$^2$,
        M. Centelles$^1$, and X. Vi\~nas$^1$}
                                                        
\affiliation{
$^1${\it Departament d'Estructura i Constituents de la Mat\`eria,
     Facultat de F\'{\i}sica,
\\   Universitat de Barcelona,
     Diagonal {\sl 647}, {\sl 08028} Barcelona, Spain}
\\
$^2${\it Institute of Physics, Sachivalaya Marg, 
         Bhubaneswar {\sl 751 005}, India.}
}
%

\begin{abstract}
We show that in the search of a unified mean field description of
finite nuclei and of nuclear and neutron matter even at high
densities, the relativistic nuclear model derived from effective field
theory and density functional theory methods constitutes a competitive
framework. The model predicts a soft equation of state, owing to the
additional meson interaction terms, consistently with the results of
the microscopic Dirac-Brueckner-Hartree-Fock theory and recent
experimental observations from heavy ion collisions. In finite
systems, after inclusion of the pairing correlations, the model is
able to describe both stable and exotic nuclei. We address two
examples at the limits of the nuclear landscape. On the one hand, we
analyze the giant halo effect and the isoscalar giant monopole
resonance in very neutron-rich Zr isotopes. On the other hand, we
discuss the structure of superheavy nuclei with double shell closures.
\end{abstract}

\maketitle

\newpage

\section{Introduction}

The mean field treatment of the relativistic field theory of hadrons
known as quantum hadrodynamics (QHD) has been found to be a very
successful framework for describing diverse bulk and single-particle
properties of nuclear matter and finite nuclei
\cite{ser86,rin96,ser97}. The QHD theory is based on a Lagrangian
density which uses the nucleon (as Dirac particle) and the
isoscalar-scalar $\sigma$, the isoscalar-vector $\omega$,
isovector-vector $\rho$ and the pseudo-scalar $\pi$ mesons as the
relevant degrees of freedom in order to understand many aspects of the
nuclear many-body problem. In the relativistic mean field (RMF)
approach the nucleus is described in terms of Dirac quasiparticles
moving in classical meson fields. Due to the definite ground-state
spin and parity of the nuclear system, the contribution of the $\pi$
meson vanishes at the mean field level. The mean field approximation
of QHD automatically generates important ingredients of the nuclear
problem like the spin-orbit force, or the finite range and density
dependence of the nuclear interaction.

The original linear $\sigma-\omega$ version of QHD \cite{ser86,wal74}
gives a very stiff equation of state (EOS) with a nuclear matter
incompressibility modulus $K_\infty \sim 600$ MeV\@. Also, the model
does not yield the average properties of the ground-state of finite
nuclei in good enough agreement with the experimental values. To
remove these deficiencies, the linear $\sigma-\omega$ model was
complemented with cubic and quartic non-linearities of the $\sigma$
meson \cite{bog77}. Adjusting some coupling contants and meson masses
from the properties of a small number of finite nuclei, the non-linear
$\sigma-\omega$ model (called hereafter standard RMF model) produces
an excellent description of binding energies, radii, spectra, and
excitation properties of spherical and deformed nuclei along the whole
periodic table \cite{rin96}. The RMF theory was proposed to be
renormalizable, and thus the scalar self-interactions were limited to
a quartic polynomial and scalar-vector or vector-vector
self-interactions were not allowed. However, the renormalizability of
the Lagrangian gets compromised by the use of coupling constants that
are not assigned with their bare experimental values but with
effective values to have proper results for finite nuclei.

More recently, inspired by the modern concepts of effective field
theory (EFT) and of density functional theory (DFT) for hadrons,
Furnstahl, Serot and Tang \cite{fur96,fur98} abandoned renormalizability and
proposed a chiral effective Lagrangian to derive an energy functional
for the nuclear many-body problem. The mean field treatment of the new
formulation, now onwards known as \mbox{E-RMF} theory, extends the
standard  RMF theory by allowing more general couplings
\cite{ser97,fur98}. An EFT assumes that there exist natural scales to
a given problem and that the only degree of freedom relevant for its
description are those which can resolve the dynamics at such scale.
The unknown dynamics, which corresponds to heavier degrees of freedom
is integrated and appears as coupling constants of the theory, which
are fitted to the known experimental data. The Lagrangian of
Furnstahl, Serot and Tang has to be understood as an EFT of low-energy
QCD. Hence, it contains the lowest-lying hadronic degree of freedom
and incorporates all the infinite couplings in general
non-renormalizable, consistent with the underlying symmetry of QCD.
Therefore, it is mandatory to develop a suitable scheme of expansion
and truncation. To do that, one assign first an index $\nu$ to each
term of the Lagrangian. This index is provided by some organizing
principle, as for instance, the naive dimensional analysis
\cite{geo93}. Next the Lagrangian is organized in powers of $\nu$ and
truncated. For the truncation to be consistent, the coupling constants
have to exhibit naturalness (i.e. they are of the order of unity when
written in an appropriate dimensionless form) and none of them can be
arbitrarily  dropped out to given order without any additional
symmetry arguments. In the nuclear structure problem, the basic
expansion parameters are the ratios of the scalar and vector fields
and of the Fermi momentum $k_F$ to the nucleon mass $M$, as these
ratios are small in normal situations (typically $k_F/M=1/3$ at
saturation  density).

From the truncated EFT Lagrangian, the energy functional can be
constructed in terms of the nuclear densities and auxilary meson
fields. Thus RMF theory can be regarded as a covariant formulation of
DFT in the sense of Kohn and Sham \cite{koh65}. From this energy
functional one can define a set of Kohn-Sham equations which minimize
the energy with respect to densities and fields. In this way all the
source terms in the Kohn-Sham equations are local. As far as the free
parameters of the RMF have been fitted to experimental data, the
corresponding mean field energy functional is a good approximation to
the exact one, unknown energy functional of the ground state densities of
the nucleonic system which includes all higher order correlations,
through the powers of the auxiliary classical mean fields. This
combination of EFT and DFT provides an approach for dealing with
the nuclear problem through Kohn-Sham (Hartree) equations with the
advantage that further contributions, at the mean field level
or beyond, can be incorporated in a systematic and controlled manner.

References \cite{fur98,fur00,rus97} have shown that it suffices to go
to fourth order in the expansion parameter $\nu$. At this level one
recovers the standard non-linear $\sigma-\omega$ plus some additional
non-linear scalar-vector and vector-vector meson interactions besides
tensor couplings. The free parameters of the energy functional have
been optimized by fitting to the ground-state observables of a few
doubly-magic nuclei, as is typical in the RMF strategy. The
corresponding fits, that were named G1 and G2 in Ref.\ \cite{fur98},
display naturalness and the results are not dominated by the last
terms retained. This confirms the usefulness of the EFT concepts and
validates the truncation of the effective Lagrangian at the first
lower orders. The ideas of EFT have allowed \cite{fur00} to elucidate
the empirical success of the previous RMF models, like the original
$\sigma-\omega$ model of Walecka \cite{wal74} and its extensions
including cubic and quartic scalar self-interactions \cite{bog77}.
However, these conventional RMF models truncate the Lagrangian at the
same level without further physical rationale or symmetry arguments.

The impact of each one of the new couplings introduced in the E-RMF
model on the properties of nuclear matter and of the nuclear surface
has been analyzed in Ref.\ \cite{est99}. That the model derived from
EFT can provide a unified framework to accomodate the successful
phenomenology of the traditional RMF models for finite nuclei and to
extend them reliably for applications in regions of higher densities
than around saturation was shown in Refs.\ \cite{est01,cen02}. In
fact, nowadays heavy ion collisions can compress nuclear matter in the
laboratory to several times the saturation density value. These large
densities are envisaged to be present within core-collapse supernovae
and their remnant neutron stars. A common practice in nuclear
phenomenology is to extrapolate the models adjusted to the
experimental properties of nuclei at normal densities for predictions
of such very dense systems. While all the models agree similarly for
the normal stable nuclei, their extrapolations can differ
significantly. It is therefore meaningful to compare the scenarios for
dense matter predicted by the RMF and E-RMF models. Microscopic
Dirac-Brueckner-Hartree-Fock calculations \cite{bro90,li92} and recent
experimental data \cite{stu01,dan02} suggest that the nuclear equation
of state (EOS) at high densities is rather soft. We will see that the
E-RMF energy functional allows one to explain the situation.

During the last decades much effort has been devoted to measure masses
of nuclei far from stability. The planned radioactive ion beam
facilities bring renewed and well-founded expectations to the field.
The available new experimental data turn out to be a demanding
benchmark for the predictions of currently existing relativistic and
non-relativistic nuclear force parameters. In Ref.\ \cite{est01a} we
verified the ability of the E-RMF model for describing the
ground-state properties of different isotopic and isotonic chains from
the valley of $\beta$ stability up to the drip lines. The residual
pairing interaction was treated in a modified BCS approach which takes
into account quasibound levels owing to their centrifugal barrier.
Also, the convergence of the E-RMF approach was studied for some
specific doubly-magic nuclei far from stability in Ref.\ \cite{hue02}
and remarkable agreement with experiment was found. In the mass region
of superheavy nuclei the E-RMF parametrizations have been applied for
investigating the next possible shell closures beyond $Z=82$ and
$N=126$, through the analysis of several indicators such as
two-nucleon separation energies, two-nucleon shell gaps, average
pairing gaps, and the shell correction energy \cite{sil04}. Other
applications of the model based on EFT include the description of
asymmetric nuclear matter at finite temperature \cite{wan00},
calculations of the Landau parameters \cite{cai01}, investigations of
the nuclear spin-orbit force \cite{fur98a}, or studies of pion-nucleus
scattering \cite{cla98}.

The rest of the paper is organized as follows. In the second section
we summarize the E-RMF model based on EFT\@. The third section is
devoted to the study of the infinite nuclear and neutron matter in the
E-RMF model, paying special attention to the possible extrapolations
to dense systems by comparing with microscopic
Dirac-Brueckner-Hartree-Fock calculations \cite{bro90,li92}. The
fourth section exemplifies the ability of the E-RMF model for
describing exotic nuclei far from the $\beta-$stability line in the
case of very neutron-rich Zr isotopes. In the fifth section we analyze
the application of the E-RMF model to superheavy elements. Often, we
will compare the predictions obtained from the parametrizations based
on EFT with the results from the standard non-linear $\sigma-\omega$
NL3 parameter set \cite{lal97}, taken as a reference. NL3 is regarded
as one of the best representatives of the RMF model with only scalar
self-interactions because of its proven performance in describing many
nuclear phenomena. A summary and the conclusions are given in the last
section.


\section{Formalism}

The E-RMF model used here has been developed in Ref.\ \cite{fur98},
where the reader can find the details of the construction of the
effective Lagrangian with a non-linear realization of chiral symmetry.
Further insight into the model and the underlying concepts can be
gained from Refs.\ \cite{ser97,fur96,fur00,rus97}. To solve the
equations of motion of the theory one applies the relativistic mean
field or Hartree approximation. The meson fields are replaced with
their ground-state expectation values, and thus they are treated as
classical fields. In such an approach the pseudoscalar field of the
pions does not contribute explicitly because it has a vanishing
expectation value. The quantum structure is introduced by expanding
the nucleon field on a single-particle basis. For systems with time
reversal symmetry, as there can be no currents, only the time-like
component of the vector meson and photon fields contributes. Charge
conservation implies that only the third component in isospin space of
the isovector rho-meson field does not vanish. As a final product, one
obtains the following energy density functional of the EFT
relativistic model for applications to finite nuclei in mean field
approach \cite{ser97,fur98}:
%
\begin{eqnarray}
\label{EDF}
{\cal E}({\bf r}) & = &  \sum_\alpha \varphi_\alpha^\dagger
\Bigg\{ -i \mbox{\boldmath$\alpha$} \!\cdot\! \mbox{\boldmath$\nabla$}
+ \beta (M - \Phi) + W
+ \frac{1}{2}\tau_3 R
+ \frac{1+\tau_3}{2} A
\nonumber \\[3mm]
& &
- \frac{i}{2M} \beta \mbox{\boldmath$\alpha$}\!\cdot\!
\left( f_v \mbox{\boldmath$\nabla$} W
+ \frac{1}{2}f_\rho\tau_3 \mbox{\boldmath$\nabla$} R
+ \lambda \mbox{\boldmath$\nabla$} A \right)
+ \frac{1}{2M^2}\left (\beta_s + \beta_v \tau_3 \right ) \Delta
A \Bigg\} \varphi_\alpha
\nonumber \\[3mm]
& & \null
+ \left ( \frac{1}{2}
+ \frac{\kappa_3}{3!}\frac{\Phi}{M}
+ \frac{\kappa_4}{4!}\frac{\Phi^2}{M^2}\right )
\frac{m_{s}^2}{g_{s}^2} \Phi^2  -
\frac{\zeta_0}{4!} \frac{1}{ g_{v}^2 } W^4
\nonumber \\[3mm]
& & \null + \frac{1}{2g_{s}^2}\left( 1 +
\alpha_1\frac{\Phi}{M}\right) \left(
\mbox{\boldmath $\nabla$}\Phi\right)^2
- \frac{1}{2g_{v}^2}\left( 1 +\alpha_2\frac{\Phi}{M}\right)
\left( \mbox{\boldmath $\nabla$} W  \right)^2
\nonumber \\[3mm]
& &  \null - \frac{1}{2}\left(1 + \eta_1 \frac{\Phi}{M} +
\frac{\eta_2}{2} \frac{\Phi^2 }{M^2} \right)
\frac{{m_{v}}^2}{{g_{v}}^2} W^2
- \frac{1}{2g_\rho^2} \left( \mbox{\boldmath $\nabla$} R\right)^2
- \frac{1}{2} \left( 1 + \eta_\rho \frac{\Phi}{M} \right)
\frac{m_\rho^2}{g_\rho^2} R^2
\nonumber \\[3mm]
& & \null
- \frac{1}{2e^2}\left( \mbox{\boldmath $\nabla$} A\right)^2
+ \frac{1}{3g_\gamma g_{v}}A \Delta W
+ \frac{1}{g_\gamma g_\rho}A \Delta R.
\end{eqnarray}
In the present form the coupling constants should be of order unity
according to the naturalness assumption. In Eq.\ (\ref{EDF}) $\tau_3$
is the third component of the isospin operator. The index $\alpha$
runs over all occupied nucleon states $\varphi_\alpha ({\bf r})$ of
the positive energy spectrum. The meson fields are $\Phi \equiv g_{s}
\phi_0({\bf r})$ (isoscalar scalar $\sigma$ meson), $ W \equiv g_{v}
V_0({\bf r})$ (isoscalar vector $\omega$ meson), and $R \equiv
g_{\rho}b_0({\bf r})$ (isovector vector $\rho$ meson), and the photon
field is $A \equiv e A_0({\bf r})$.

From the functional (\ref{EDF}) one derives the field equations obeyed
by the nucleonic and the mesonic fields. The variation of Eq.\
(\ref{EDF}) with respect to $\varphi^\dagger_\alpha$ gives the Dirac
equation fulfilled by the nucleons, and the variations with respect to
the various meson fields result in the Klein-Gordon equations obeyed
by the mesons. The expressions for the densities and the field
equations of the E-RMF model can be found in previous works
\cite{ser97,fur98,est01,cen02,hue02} and we shall not repeat them
here. In practice we solve the Dirac equation in coordinate space by
transforming it into a Schr\"odinger-like equation and iterate
numerically the final set of coupled equations till consistency is
reached. 

Let us now describe the more relevant implications of the new terms of
the energy density (\ref{EDF}). With the inclusion of the quartic
vector self-interaction $\zeta_0$ and the terms $\eta_1$ and $\eta_2$
one is able to obtain a desirable positive value of the coupling
constant $\kappa_4$ of the quartic scalar self-interaction. This is so
for realistic nuclear matter properties and at the same time keeping
all the parameters within the bounds of naturalness. Furthemore, these
bulk couplings $\zeta_0$, $\eta_1$, and $\eta_2$ confer an extra
density dependence to the scalar and vector self-energies
\cite{est01,cen02} which is consistent with the output of microscopic
Dirac-Brueckner-Hartree-Fock calculations that start from the bare
nucleon-nucleon interaction in free space. In the E-RMF model the bulk
symmetry energy coefficient $J$ depends on the coupling $\eta_\rho$ in
addition to the usual coupling $g_\rho$. The new coefficient
$\eta_\rho$ is useful to fit $J$ and in turn to tune, relative to $J$,
the stiffness of the nucleus against pulling neutrons apart from
protons as the neutron excess is increased \cite{est99}.

In the standard RMF model the only parameter not related with the
saturation conditions is the mass of the scalar meson $m_s$. The
additional couplings $\alpha_1$ and $\alpha_2$ in the E-RMF energy
density are helpful to improve on the description of the nuclear
surface properties (e.g., surface energy and surface thickness)
without spoiling the bulk properties \cite{est99}. The quantities
$\beta_{s}$, $\beta_{v}$, $g_\gamma$ and $\lambda = \frac{1}{2}
\lambda_{p} (1 + \tau_3) + \frac{1}{2} \lambda_{n}(1 - \tau_3)$ take
care of effects related with the electromagnetic structure of the pion
and the nucleon (the constants $g_\gamma^2/4\pi = 2.0$, $\lambda_{p} =
1.793$ and $\lambda_{n}=-1.913$ are given their experimental values)
\cite{fur98}. The tensor coupling $f_v$ between the $\omega$ meson and
the nucleon adds momentum and spin dependence to the interaction. It
introduces a corrective term in the spin-orbit potential as compared
with the expression in the standard RMF model \cite{est01a,fur98a}.
Due to the existence of a trade-off between the size of the $\omega$
tensor coupling and the size of the scalar field, it is possible to
obtain parameter sets that provide excellent fits to nuclear masses,
radii and spin-orbit splittings with a larger value of the equilibrium
effective mass than in models that ignore such coupling.

In the applications to be presented below we shall employ the E-RMF
parameter sets G1 and G2 of Ref.\ \cite{fur98}. The masses of the
nucleon and of the $\omega$ and $\rho$ mesons are $M= 939$ MeV,
$m_{v}= 782$ MeV, and $m_\rho= 770$ MeV, respectively. The parameters
$m_{s}$, $g_{s}$, $g_{v}$, $g_\rho$, $\eta_1$, $\eta_2$, $\eta_\rho$,
$\kappa_3$, $\kappa_4$, $\zeta_0$, $f_{v}$, $\alpha_1$, and $\alpha_2$
of G1 and G2 were fitted by a least-squares optimization procedure to
29 observables (binding energies, charge form factors and spin-orbit
splittings near the Fermi surface) of the nuclei $^{16}$O, $^{40}$Ca,
$^{48}$Ca, $^{88}$Sr and $^{208}$Pb, as described in Ref.\
\cite{fur98}. The constants $\beta_{s}$, $\beta_{v}$ and $f_\rho$ were
then chosen to reproduce the experimental charge radii of the nucleon.
We report in Table~1 the values of the parameters and the saturation
properties of the sets G1 and G2 as well as those of the NL3
parameters \cite{lal97}. An interesting feature is that the set G2 has
a positive $\kappa_4$ coupling, as opposed to G1 and to most of the
successful RMF parametrizations such as NL3. Formally, a negative
value of $\kappa_4$ is not acceptable because the energy spectrum then
has no lower bound \cite{bay60}. We note that the value of the
effective mass at saturation $M^*_\infty/M$ in the EFT sets ($\sim
0.65$) is larger than the usual value in the RMF parameter sets ($\sim
0.60$), which is due to the presence of the tensor coupling $f_{v}$ of
the $\omega$ meson to the nucleon. Also, the nuclear matter
incompressibility of the G1 and G2 sets ($K=215$ MeV) is visibly
smaller than that of the NL3 set ($K=271$ MeV).

\section{Nuclear and neutron matter}

In an infinite medium of uniform nuclear matter all of the terms with
gradients  in the energy density ${\cal E}$ and in the field equations
vanish. In this limit the nucleon density is given by
\begin{equation}
\rho= \frac{\gamma}{(2\pi)^3} \int_0^{k_F} d^3 k =
\frac{\gamma}{6{\pi^2}}{k_F^3} ,
\end{equation}
where $k_F$ is the Fermi momentum, and the degeneracy factor $\gamma$
is 4 for symmetric nuclear matter and 2 for pure neutron matter. The
reader can find the E-RMF expression of ${\cal E}$ for asymmetric
nuclear matter at finite temperature in Ref.\ \cite{wan00}.

The relevance of relativistic effects in the nuclear EOS was soon
realized when the relativistic Dirac-Brueckner-Hartree-Fock (DBHF)
calculations provided a clue for solving the Coester band problem
\cite{bro90}. The microscopic DBHF theory suggests a soft EOS at high
densities \cite{bro90,sug94}. The recent experimental data
\cite{stu01,dan02} also rule out the possibility of a strongly
repulsive nuclear EOS\@. Typical representatives of the standard RMF
theory like the NL3 parameter set cannot follow the trend of the DBHF
results even at slightly high densities \cite{est01}. In contrast to
the conventional RMF model, the E-RMF calculations at high density
regimes yield results in accordance with the DBHF theory. It has to be
noted that relativistic models which resort to density-dependent
couplings are also consistent with the DBHF calculations \cite{typ03}.

In Figure 1 we present the density dependence of the nuclear matter
scalar and vector self-energies calculated with G1, G2, NL1
\cite{rei86} and NL3 versus the DBHF result. While G2 follows the
nature of the DBHF self-energies quite remarkably, at densities only
slightly above saturation the NL3 results soon depart from the DBHF
behaviour. Thus, the success of the usual RMF model with only scalar
self-interactions for describing the saturation point and the data for
finite nuclei is not followed by a proper description of the
microscopic DBHF calculations. This is caused importantly by a too
restrictive treatment of the $\omega$-meson term
\cite{est01,cen02,sug94}. While in the standard RMF model the vector
potential increases linearly with density and gets stronger, in DBHF
it bends down with density. Moreover, the scalar potential
overestimates the DBHF result at high density in order to compensate
for the strong repulsion in the vector channel. The additional self-
and cross-interactions $\zeta_0$, $\eta_1$, and $\eta_2$ included in
the E-RMF sets result in a richer density dependence of the mesonic
mean fields which brings about the improvement in comparison with the
DBHF calculations \cite{est01}.

Figure 1 demonstrates the importance of meson self-interactions at
higher densities and exposes the inadequacy of restricted models which
neglect them for applications to such conditions. This argument is
further supported by Figure 2 in which the variation of the binding
energy per particle is plotted as a function of the density. We can
see that the calculations of dense matter based on the RMF sets NL1
and NL3 deviate largely from DBHF, while the E-RMF calculations with
G1 and, specially, G2 agree better with the density dependence of the
EOS of the DBHF theory. One can realize that in spite of the fact that
the incompressibility of NL1 is within the empirical boundaries
($K_{\infty}=212$ MeV for NL1), the EOS of this set soon becomes stiff
with increasing density and does not follow the DBHF trend. The E-RMF
parameter sets give a soft EOS both around saturation and at high
densities. A similar situation prevails in the EOS of neutron matter
(Figure 3), though the agreement of G1 and G2 with the DBHF
calculation is not as remarkable as in the case of symmetric matter.
From the point of view of the comparison with DBHF, this would
indicate that the present E-RMF model still needs an improvement in
the treatment of the isovector sector, like consideration of
additional cross couplings involving the rho-meson field or the
introduction of an isovector scalar meson.

The average densities of terrestrial nuclei are not very far from the
values around the saturation point of the nuclear EOS, where the E-RMF
sets produce similar results to NL3. Hence, one can expect that in
finite nuclei the G1 and G2 interactions also will yield results on a
par with the celebrated NL3 parametrization. That this is indeed the
case is illustrated in the next section for the bulk properties of
finite nuclei near and away from the $\beta$-stability valley.

\section{Structure of exotic nuclei}

Exotic nuclei far from the stability line have attracted much
attention in the nuclear physics community from both the theoretical
and the experimental sides. One expects to find very different
properties from the normal nuclei as soon as one leaves the
$\beta$-stability region and approaches the drip lines. Neutron-rich
nuclei near the drip line and the occurrence of closed shells are very
important in nuclear astrophysics because their properties strongly
influence how stable neutron-rich nuclei are formed through the
r-process. It is expected that in very neutron-rich nuclei the shell
structure be strongly modified, with some of the traditional shell
gaps disappearing and with new ones appearing.

Another interesting feature of some exotic nuclei is the appearence of
a halo structure which was experimentally discovered in $^{11}$Li
\cite{tan85} and which has also been observed in $^{11,14}$Be and
$^{17}$B \cite{tan96}. In some heavier neutron rich nuclei a sudden
increase of the neutron radii close to the neutron drip line, the
so-called giant halo, has been predicted by theoretical relativistic
Hartree-Bogoliubov (RHB) calculations for Zr \cite{men98} and Ca
\cite{men90} isotopes.

Neutron-rich nuclei near the drip line usually have a very small Fermi
level and thus valence nucleons can easily scatter to the continuum
states through the pairing correlations. Consequently, it is mandatory
to properly take into account the coupling between the bound and the
continuum for dealing with nuclei near the drip lines. In the
relativistic domain, the microscopic HFB theory should, in principle,
be used. Although the simple BCS theory fails in describing nuclei
near the drip lines \cite{dob84}, it can still be used if some
refinements to the standard BCS method are added, such as the resonant
continuum coupling \cite{san00,san03} or if one takes into account
quasibound levels owing to their centrifugal barrier \cite{est01}.
These variations of the BCS method in general allow one to describe
nuclei in the vicinity of the drip lines with very reasonable accuracy
and avoiding the difficulties of a full RHB calculation.

In this section, we analyze the giant halo effect in Zr isotopes
using  the E-RMF approximation together with the pairing prescription
of  Ref.\ \cite{est01} which allows one to describe isotopic
(isotonic)  chains with magic $Z$ ($N$) numbers from the proton to the
neutron  drip lines. As far as many properties of the exotic nuclei
are  modified when one approches the drip line, we will discuss
afterwards  a different problem related with excited states, namely,
the excitaion energy of the isoscalar giant monopole resonance. From
nonrelativistic  RPA calculations \cite{ham97} it is known that in
nuclei near the drip line the monopole strength distribution is much
affected by the presence of the low-energy threshold stemming from
tiny bound nucleons. We study here whether the E-RMF parametrizations
in  constrained calculations are able to identify the main trends
exhibited by the monopole RPA strength near the drip lines.

\subsection{Treatment of pairing}

To deal with the pairing correlations we shall use here a simplified
prescription which has proven to be in good agreement with RHB
calculations \cite{est01a}. For each kind of nucleon we assume a
constant pairing matrix element $G_q$, which simulates the zero range
of the pairing force, and include quasibound levels in the BCS
calculation as done in Ref.\ \cite{cha98}. These quasibound states mock
up the influence of the continuum in the pairing calculation. We
restrict the available space of $\alpha_q$ states to one harmonic
oscillator shell above and below of the Fermi level, to avoid the
unrealistic pairing of highly excited states and to confine the region
of influence of the pairing potential to the vicinity of the Fermi
level. As explained in Ref.\ \cite{est01}, the solution of the pairing
equations provides the chemical potential $\mu_q$ and the average
pairing gap $\Delta_q$ for each kind of nucleon.

The quasibound levels of positive single-particle energy are retained
by the centrifugal barrier (neutrons) or the centrifugal-plus-Coulomb
barrier (protons). The wave functions of the considered quasibound
levels are mainly localized in the classically allowed region and
decrease exponentially outside it. As a consequence, the unphysical
nucleon gas which surrounds the nucleus if continuum levels are
included in the normal BCS approach is eliminated. We have shown in
Ref.\ \cite{est01a}, by comparison with available RHB results, that the
procedure is able to predict well the position of the proton and
neutron drip lines or, e.g., the behavior of the neutron and charge
radii far from stability. Also, the calculated pairing gaps turn out
to be scattered around the empirical average $12/\sqrt{A}$ MeV\@.

\subsection{Giant halo in Zr isotopes}

As we have mentioned, giant halo nuclei are predicted by the RMF
calculations near the neutron drip line for Zr \cite{men98} and Ca
\cite{men90} isotopes. With the new radioactive isotope beam
facilities it is expected to reach experimentally the lighter giant
halo nuclei of Zr. Our calculation is performed with the E-RMF set G2
in spherical symmetry due to the fact that non-relativistic and
relativistic calculations above $^{122}$Zr predict the nuclei to be
spherical \cite{men98}. The neutron and proton constant matrix
elements are parametrized as $G_q=C_q/A$ ($q=p,n$). In the
calculations presented in this section we choose $C_n=16$ MeV and
$C_p=14$ MeV because with these pairing constants we obtain pairing
correlation energies (i.e., binding energies referred to the E-RMF
values without pairing) from $^{122}$Zr to $^{138}$Zr similar to those
displayed in Figure 5 of Ref.\ \cite{san03}. The latter calculation
was performed in the BCS-plus-resonant (r-BCS) continuum approximation
with the RMF parameter set NL-SH employing a zero-range pairing force,
which simulated well the RHB calculations of Ref.\ \cite{men98} with
NL-SH. 

One of the signatures of the giant halo consists in a sudden increase
of the neutron root mean square (rms) radius in the isotopic chain due
to the scattering of Cooper pairs to the continuum containing
low-lying resonances of small angular momentum, which in our approach
are represented by the quasibound levels. Figure 4 displays the
neutron and proton rms radii for the Zr isotopic chain from $A=80$ to
$A=120$ obtained using our pairing approach with the G2 parameter set.
One can clearly appreciate the kink at $A=122$. The neutron radii
obtained with G2 reproduce the overall trends of the results of Ref.\
\cite{san03} that were computed with the NL-SH set in the r-BCS
approximation. It is to be noted that we use a pairing force with a
constant matrix element instead of a zero-range pairing force as in
Ref.\ \cite{san03}. Another difference with Ref.\ \cite{san03} is that
our calculation does not take into account the resonance width, which
may contribute to reduce the pairing gap near the drip line.

The Zr isotopes with a neutron number beyond $N=82$ develop a large
shell gap and start progressively to fill up the weakly bound and
quasibound (continuum) $2f_{7/2}$, $3p_{3/2}$, $3p_{1/2}$, $2f_{5/2}$,
$1h_{9/2}$ and $1i_{13/2}$ single-particle neutron energy levels. This
can be seen in Figure 5 where the neutron spectra of two
representative nuclei, $^{128}$Zr and $^{140}$Zr, are displayed. In
our model the $1i_{13/2}$ and $1h_{9/2}$ are quasibound levels and
correspond to low-lying resonant levels \cite{san03}. The $3p_{1/2}$
and $2f_{5/2}$ levels lie in the continuum up to $A=134$ and $A=140$,
respectively, while the $2f_{7/2}$ and $3p_{3/2}$ states are always
bound levels from $A=124$ on. It should be pointed out that the
considered levels lying in the continuum are always quasibound levels
for $N=82$ owing to their relatively high centrifugal barrier ($l>2$).
In our calculation only the $3p_{1/2}$ orbit does not appear as a
quasibound level because of its small centrifugal barrier.


It is just the occupancies of $2f_{7/2}$, $3p_{3/2}$, $3p_{1/2}$, and
$2f_{5/2}$ levels which mainly contribute to the enhancement of the
neutron rms beyond $N=82$, as discussed in earlier literature
\cite{men98,san03}. In the lighter Zr isotopes one observes the
progressive filling of the $1h_{11/2}$ state and the almost negligible
occupancy of the states beyond the $N=82$ core. After $N=82$, the
occupancy of the weakly bound and quasibound $3p_{3/2}$, $3p_{1/2}$,
$2f_{5/2}$, $1h_{9/2}$ and $1i_{13/2}$ states starts to be
progressively important when the number of neutrons beyond the core
increases. In our model, the $3p_{1/2}$ state appears when $A=134$ as
a tiny bound state which causes the small kink between $A=132$
and $A=134$ that can be seen in Figure 4.

The rms radii of the neutron orbits above the $N=82$ core and
of the $2d_{3/2}$ and $1h_{11/2}$ states  of the
core, as well as the radii of the quasibound $1h_{9/2}$
and $1i_{13/2}$ levels, are displayed in Figure 6. The large
rms radii of the $2f_{5/2}$, $3p_{3/2}$ and $3p_{1/2}$ orbits can
clearly be seen in this figure. These large rms radii are basically
due to two different reasons. First, the principal quantum number $n$
is large (2 or 3) which makes the wave functions corresponding to
these orbitals extend quite far. The second reason lies in the fact
that the $2f_{5/2}$, $3p_{3/2}$ and $3p_{1/2}$ levels are extremely
weakly bound, or even unbound, for $N>82$. For instance, the single
particle energy of the $3p_{3/2}$ level runs from $-$0.16 MeV in
$^{124}$Zr to $-$0.77 MeV in $^{140}$Zr. Due to the last reason, these
levels show up a relatively strong dependence with the mass number
$A$. However, the rms radii of the $1h_{11/2}$ and $2f_{7/2}$ orbitals
are roughly constant with $A$ due to the larger binding energy of
these states. The quasibound $1h_{9/2}$ and $1i_{13/2}$ states also
exhibit a nearly constant rms radii as a function of $A$. In this case
the reason is rather that the small principal quantum number ($n=1$)
pushes the significant part of the wave function inside enough for not
being affected by the increasing number of nucleons outside the
$^{122}$Zr core. We note that for obtaining the contribution of the
states displayed in Figure 6 to the total neutron rms radius, the
degeneracy $2j+1$ and the occupancy $v^2$ of these levels has to be
taken into account, which gives an increasing contribution with $A$.
For instance, the contribution to the total $\langle r_n^2 \rangle$
radius from the $2f_{7/2}$, $3p_{3/2}$ and $2f_{5/2}$ states is 0.59,
0.50 and 0.06 fm$^2$ in $^{124}$Zr, whereas it is 2.63, 1.80 and 1.53
fm$^2$ in the drip line nucleus $^{140}$Zr. This explains the
increasing tendency of the rms radii in Figure 4 beyond the
$N=82$ core.

Another indicator of a giant halo is a very small neutron chemical
potential $\mu_n$. This is just the situation we find using the E-RMF
set G2, with the discussed pairing prescription, in the neutron rich
Zr isotopes beyond $N=82$. In our calculation, $\mu_n$ ranges from
$-$0.89 MeV for $^{124}$Zr to practically zero in $^{140}$Zr. As a
consequence of this almost vanishing value of $\mu_n$ beyond the
$N=82$ core, the binding energies are roughly constant, ranging from
943.7 MeV in $^{124}$Zr to 951.9 MeV in $^{140}$Zr, and the
two-neutron seperation energies $S_{2n}$ are close to zero. We have
calculated the two-neutron separation energies along the Zr isotopic
chain for the G2 and NL-SH parameter sets. In both parameter sets
shell effects appear at $N=50$ and 82, as expected. Beyond $N=82$,
$S_{2n}$ decreases faster for NL-SH than for G2 indicating that
although the general features of giant halo nuclei are unchanged, fine
details can be force dependent. In Figure 7 we display the calculated
neutron densities of some selected Zr isotopes. When the neutron
number is significantly above the $N=82$ core the central density
starts to progressively decrease, while a halo develops in the outer
region. Some neutrons of the core are scattered by the pairing
correlations to the states beyond the Fermi level, with larger rms
radii with the increasing number of the neutrons in the outer part of
the density distribution of the nuclei beyond $^{122}$Zr.

The simplified treatment of the pairing correlations used here, which
is discussed in more detail in Ref.\ \cite{est01a}, is thus seen to be
able to describe the bulk properties of nuclei near the drip lines
\cite{est01,est01a} and also to reproduce correctly the main features
of the giant halo nuclei. Only some finer details such as the behavior
of the $3p$ levels in the giant halo problem discussed here, would
require of an improved treatment of the pairing correlations using
more sophisticated methods such as the r-BCS or RHB approaches. On the
other hand, the study presented here, together with the examples
discussed in Refs.\ \cite{est01,est01a}, proves the ability of the RMF
approach based on EFT for describing exotic isospin-rich nuclei far
from the $\beta$-stability valley when a pairing residual interation
is included. Notice that only information about a few stable magic
nuclei was used in the fit of the coupling constants of the E-RMF
sets, and thus the results near the drip lines are predictions of the
model.

\subsection{Breathing mode energy near the neutron drip line}

It is to be expected that in neutron rich nuclei, and in particular in
giant halo nuclei, the excitation energy of collective modes be
strongly modified as compared with stable nuclei. For further
prospect, we next address with the example of the Zr isotopes the
trends predicted by the G2 set for the behavior of the excitation
energy of the isoscalar giant monopole resonance (GMR) in approaching
the limits of neutron stability. As we have seen in the previous
subsection, in Zr isotopes beyond $A=122$ the last filled levels are
weakly bound. Thus, particle-hole excitations from the occupied
$2f_{7/2}$, $3p_{3/2}$, $3p_{1/2}$ and $2f_{5/2}$ states to the
continuum will give important contributions to the low-energy part of
the RPA strength. The situation for these Zr isotopes is similar to
that found in the $^{110}$Ni nucleus for which the RPA strength has
been analyzed in Ref.\ \cite{ham97} by means of non-relativistic
Skyrme force calculations. Although in such kind of very neutron-rich
nuclei the monopole RPA strength is broad and it is not concentrated
in a single peak \cite{ham97}, the mean energy of the monopole
response should exhibit a downwards behavior when one moves from
stable nuclei towards the neutron drip line due to the increase of the
amount of strength in the low energy region of the resonance.

Constrained calculations are one of the means to estimate the average
energy of the breathing mode. For example, RMF constrained
calculations have been performed with different parameter sets to
obtain the excitation energies of the GMR of some selected nuclei in
Refs.\ \cite{mar89,sto94,sto94a,pat02}. A reasonable good agreement
has been found between the results of these calculations with the
results from more sophisticated approaches such as the time-dependent
RMF formalism \cite{vre97} or the relativistic RPA \cite{pie01}.

Here we follow the method of Refs.\ \cite{mar89,sto94,sto94a,pat02}
to perform the constrained calculations with the E-RMF model. One has
to solve the self-consistent equations for the nucleonic and mesonic
fields derived from the constrained energy
\begin{eqnarray}
E_c(\eta)=\int d{\bf r} \left[{\cal E} -\eta \sum_\alpha
\varphi^\dagger_\alpha(r^2-R_0^2)\varphi_\alpha\right] ,
\label{GM1}
\end{eqnarray}
where ${\cal E}$ is the energy density given by Eq.\ (1) and
$R_0$ the rms radius of the ground state. The solution of the
corresponding equations of motion gives spinors $\varphi_\alpha (\eta)$
and mesonic fields which depend on the Lagrange multiplier $\eta$.

The binding energy $E(\eta)$, obtained by integrating Eq.\ (1) with
the nuclear and meson fields solution to the constrained problem, has
a minimum at $\eta=0$ which corresponds to the ground-state energy
with rms radius $R_0$. Expanding $E(\eta)$ in a harmonic approximation
about $R_0$ one obtains the constrained incompressibility of the
finite nucleus as
\begin{eqnarray}
K_A^c={\frac{1}{A}R_0^2
\frac{\partial^2 E(\eta)}{\partial R_\eta^2}\vline}_{\eta=0} ,
\label{GM2}
\end{eqnarray}
with $R_\eta$ being the rms radius computed with the constrained
spinors $\varphi_\alpha (\eta)$. The constrained incompressibility of
Eq.\ (\ref{GM2}) is just the constant entering the harmonic restoring
force. 

The mass parameter of the monopole oscillation is given by
\begin{eqnarray}
B=\frac{1}{A} \int d{\bf r} \, u({\bf r})^2 {\cal E}({\bf r}) ,
\label{GM3}
\end{eqnarray}
where $u({\bf r})$ is the displacement field which is determined by
the solution of the continuity equation for the monopole oscillation
\cite{mar89,sto94,sto94a,pat02}.
In the non-relativistic limit and with a Tassie transition density,
which assumes that there exists a single-collective state,
the displacement field is $u(r)=r$ and
the expression of the mass parameter becomes
\begin{eqnarray}
B_{nr}=M \int d{\bf r} \, r^2 \rho({\bf r})=M \langle r^2 \rangle .
 \label{GM4}
\end{eqnarray}
The frequency of the constrained monopole vibration using the
non-relativistic mass parameter is
\begin{eqnarray}
E_{c}=\sqrt{\frac{AK_A^c}{B_{nr}}} .
\label{GM5}
\end{eqnarray}
We have performed the constrained calculation switching off the
pairing correlations, as in the non-relativistic studies of giant
resonances in nuclei near the drip lines \cite{ham97}. Our estimate of
the excitation energy of the GMR of the nucleus $^{90}$Zr from the
constrained calculation with the E-RMF set G2 is 17.2 MeV\@.
It agrees fairly well with the experimental
centroid energy $17.9\pm 0.2$ MeV \cite{you99}.

Figure 8 displays the finite nucleus incompressibility $K_A^c$
calculated using the G2 set for even Zr isotopes ranging from $A=90$
to $A=138$. The finite nucleus incompressibilities are roughly
constant from $A=90$ up to $A=122$, when the $1h_{11/2}$ level is
completely filled. From $A=122$ on, the finite nucleus
incompressibility decreases quite fast till the neutron drip line is
reached. One can try to understand qualitatively this downward
tendency as follows. Inspired by the liquid drop formula, the finite
nucleus incompressibility can be expanded \cite{bla80} into a bulk
contribution $K_\infty$ (the nuclear matter incompressibility
modulus), plus surface $K_{\rm surf}$, Coulomb $K_{\rm Coul}$ and
volume-symmetry $K_{\rm sym}$ terms, apart from corrections of higher
order. Within a scaling model approach the quantities $K_{\rm surf}$,
$K_{\rm Coul}$, and $K_{\rm sym}$ are negative \cite{pat02,pat02a}.
When the mass number $A$ increases approaching the drip lines, the
surface and Coulomb terms become comparatively less important.
However, the symmetry contribution, which is large in the RMF
parametrizations, multiplied by a large average asymmetry factor
$(N-Z)^2/A^2$ makes finally the nuclei near the neutron drip line
softer than the stable nuclei. Though the numerical coefficients of
the expansion of the nucleus incompressibility $K_A$ into its various
contributions are different between the scaling and the constrained
models \cite{bla80,jen80}, one can still expect that the
volume-symmetry term is at the origin of the reduction of $K_A^c$ as
the nuclei come close to the neutron drip point.

From a more microscopic point of view, when $A>122$ the weakly bound
neutron levels $2f_{7/2}$, $3p_{3/2}$, $3p_{1/2}$ and $2f_{5/2}$,
which have a large rms radius (see Figure 6), start to be occupied.
Neutrons in these levels behave almost as extremely asymmetric nuclear
matter, which is strongly softened as compared with the symmetric one
\cite{bom91}. According to this, when the number of neutrons outside
the $A=122$ core increases, their contribution to $K_A^c$ is smaller
and thus, on average, the finite nucleus incompressibility of the Zr
isotopes that develop a giant halo is smaller than the one of the
stable Zr nuclei.

Due to the fact that the Zr nuclei beyond $A=122$ have a large rms
radius, the mass denominator (\ref{GM4}) also increases when one moves
from $A=122$ towards the neutron drip line. This fact together with
the decreasing of the finite nucleus incompressibility reduces
strongly the excitation energy of the GMR beyond $A=122$, as evidenced
by Figure 8. It is known that the experimental GMR excitation
energies roughly follow the empirical law $E_{\rm GMR}\sim 80A^{1/3}$
MeV \cite{bla80}. The excitation energy of the GMR of the Zr isotopes
from $A=90$ to $A=122$ can be nicely fitted by an $A^{1/3}$ law, as
indicated by the dashed line in Figure 8. This behaviour breaks down
for the isotopes with $A>122$ pointing out again that for these nuclei
the neutrons which occupy the weakly bound levels above the
$1h_{11/2}$ orbit are softer than the neutrons belonging to the
$A=122$ core.

As shown in Ref.\ \cite{pat02}, the energy
of the constrained monopole oscillation (\ref{GM5}) can be 
nominally written in the nonrelativistic language of RPA sum rules
as $E_c=\sqrt{m_1/m_{-1}}$
\cite{boh79}.
The sum rules, or moments of the strength function $S(E)$ of the giant
resonance, are defined as
\begin{equation}
m_k = \int^{\infty}_0 E^k S(E) dE
\label{eqmk} \end{equation}
for integer $k$. The $m_1$ sum rule turns out to be proportional to
the mass denominator $B_{nr}$, while the inverse energy weighted sum
rule $m_{-1}$ can be computed as \cite{boh79}
\begin{eqnarray}
m_{-1}= {-\frac{1}{2}A\frac{\partial R_\eta^2}{\partial 
\eta}\vline}_{\eta=0}={\frac{1}{2} \frac{\partial^2
E(\eta)}{\partial \eta^2}\vline}_{\eta=0} .
\label{GM7}
\end{eqnarray}
From this point of view, the strong reduction of ${E}_c$ when $A>122$
is basically due to an increasingly large inverse energy weighted
moment of the RPA strength. The latter, due to the power $E^{-1}$ in
(\ref{eqmk}),  reflects the important contribution to the low-energy
part of the RPA strength distribution which arises from the
particle-hole transitions to the continuum from the weakly bound
energy levels $2f_{7/2}$, $3p_{3/2}$, $3p_{1/2}$ and $2f_{5/2}$.

\section{Predictions in the land of superheavy elements}

In continuance of our study of the nuclear structure predictions of
the E-RMF model in the regions of the nuclear landscape away from the
stable nuclei where the parameters were fitted, in this section we
address spherical calculations of superheavy elements (SHE). We shall
concentrate on finding out where the next doubly-magic shell closures
beyond $N=126$ and $Z=82$ are located according to the E-RMF model.

Experimental efforts at the leading laboratories for synthesizing new
elements in the superheavy mass region have already produced some
light isotopes of $Z=110$--$112$ at GSI and Dubna
\cite{hof02} and even heavier isotopes of $Z=112$--$116$
at Dubna \cite{oga02}. Most of the detected nuclei with $Z\sim110$ are
deformed, consistently with the predicted occurrence of a deformed
shell closure at $Z=108$ and $N=162$ \cite{bur98,mol94,smo97}.
Therefore, quadrupole deformations
\cite{bur98,mol94,smo97,ren02,gor02}, and even in some cases triaxial
calculations \cite{bur04}, have to be taken into account
for describing the new nuclei around $Z=110$. However, when one is
concerned with identifying the location of the $N$ and $Z$ values of
the next spherical {\it double} shell closure beyond $^{208}$Pb, most
of the calculations with relativistic and non-relativistic mean field
models have been performed in spherical symmetry
\cite{ben99,kru00,ben2,rei02}. Of course, one has to keep in
mind that the study of the spherical shell structure is valid only for
the doubly-magic nuclei. The addition of deformation degrees of
freedom in the calculations would certainly change the picture in the
details and add deformed shell closures, e.g., like those predicted
around $Z=108$ and $N=162$ \cite{bur98,mol94,smo97}. But it should not
change drastically the predictions for the values of $N$ and $Z$ where
the strongest shell effects show up already in the spherical
calculation. Certainly, for a quantitative discussion, one needs to
account for deformation effects which would serve to extend the island
of shell stabilized superheavy nuclei and to decide on the specific
form of the ground-state shapes of those nuclei.

In the nonrelativistic framework, the macroscopic-microscopic models
predict spherical shell closures at $Z=114$ and $N=184$ \cite{mol94}. The
Hartree-Fock calculations with Skyrme forces show, in general, the
most pronounced shell effects at $Z=124$, $Z=126$ and $N=184$ 
\cite{ben99,kru00,ben2,rei02}.
However, the conventional RMF theory typically prefers $Z=120$ and
$N=172$ as the best candidates for spherical shell closures 
\cite{ben99,kru00,ben2,rei02}.
In view of these discrepancies in the predicted shell closures for
SHE, it is interesting to reinvestigate them using the more general
E-RMF model.

\subsection{Shell closures}

In normal nuclei a large gap in the single-particle spectrum is
interpreted as the indication of a shell closure. We start by
inspecting the neutron single-particle spectra of the $^{292}120$,
$^{304}120$ and $^{378}120$ nuclei, displayed in Figure 9,
calculated using the G1, G2 and NL3 parameter sets. As mentioned,
$Z=120$ is found to be a magic number in many RMF calculations
\cite{ben99,kru00,ben2,rei02}.
The three parameter sets show a large gap at $N=172$ and
$N=258$. However, for $N=184$ a moderate gap is found mainly for the
E-RMF sets, G1 and G2, which is smaller for NL3. In the level spectrum
of the system with $N=258$ and $Z=120$, one can again find appreciable
energy gaps across the neutron numbers $N=198$ ($1j_{13/2}$ level) and
$N=228$ ($1k_{17/2}$ level) in all the considered parameter sets. By
comparing the spectra for the three systems with $N=172$, 184 and 258,
it can be noticed that the gap between two particular levels is
strongly modified along the isotopic chain. However, the level gaps
are not as distinct as in lighter nuclei. In calculations of SHE the
level spectra become very involved due to the presence of levels with
a high degree of degeneracy. Therefore, it is imperative to look for
other quantities to be able to reliably identify the shell closures
and magic numbers of SHE, apart from the analysis of the
single-particle level structure.

We next shall consider three energy indicators for locating the
nucleon shell closures. First, a sudden jump in the two-neutron
(two-proton) separation energies of even-even nuclei, defined as
$S_{2q}= E(N_q-2)-E(N_q)$, where $N_q$ is the number of neutrons
(protons) in the nucleus for $q= n$ ($q= p$). Second, the two-neutron
(two-proton) shell gap defined as the second difference of the binding
energy \cite{ben99}: $\delta_{2q}(N_q)= E(N_q+2)-2E(N_q)+E(N_q-2)$.
This quantity measures the size of the step found in the two-nucleon
separation energy and, therefore, it is strongly peaked at magic shell
closures. Third, for closed shell nuclei the average pairing gaps
$\Delta_q$ obtained in the calculations should vanish.

In Figure 10, we display these energy indicators calculated with the
E-RMF set G2 and with NL3 along the isotopic chain of $Z=120$. The
two-neutron separation energies $S_{2n}$, displayed in the top panel
of the figure, show sudden jumps after the neutron numbers $N=172$,
184 and 258 indicating possible neutron shell closures. The
two-neutron shell gap $\delta_{2n}$ along the same isotopic chain is
displayed in the middle panel of Figure 10. Sharp peaks in
$\delta_{2n}$ are found at the same neutron numbers 172, 184 and 258,
though the peak at $N=184$ is less pronounced than the peaks for
$N=172$ and $N=258$. Also, the peak in $\delta_{2n}$ at $N=258$ is
more marked in NL3 than in G2. The curve for the neutron pairing gaps
displayed in the bottom panel of Figure 10 shows a structure of arches
which vanish only at $N=172$, 184 and 258 with G2, and at $N=172$ and
258 with NL3. Therefore, all the three analyzed observables point out
to the same neutron numbers as the best candidates for shell closures
for $Z=120$ with the G2 parametrization. The neutron pairing gap
calculated with NL3 does not vanish at $N=184$. The proton pairing gap
is zero along the whole isotopic chain, for both G2 and NL3, and thus
we have not plotted it. The results obtained with the set G1 show the
same global nature as with G2 and NL3. The peak in $\delta_{2n}$
occurring at $N=172$ is lower in G1 than in G2, and the peak at
$N=184$ is very much quenched in the G1 set. As in the case of NL3,
the neutron gap $\Delta_n$ at $N=184$ does not vanish for G1
($\Delta_n \approx 0.7$ MeV).

Now we proceed to the discussion of the isotonic chains of $N=172$ and
$N=184$ obtained with E-RMF set G2 and which are displayed in Figure
11. For $N=172$ all the indicators show a very robust closure at
$Z=120$ and a much weaker shell closure at $Z=114$. This is
appreciated, e.g., from the strength of the corresponding two-proton
shell gap $\delta_{2p}$. For $N=184$ proton shell closures are found
at $Z=114$ and $Z=120$. In this case the strength of $\delta_{2p}$ for
both atomic numbers is similar (though smaller than the two-proton
shell gap found with the combination $N=172$ and $Z=120$). Moreover,
from the left bottom panel of the Figure 11, it can be seen that
below $Z=110$, the neutron pairing gap does not vanish, meaning that
the shell closure for $N=172$, which is very strong combined with
$Z=120$, is washed out. Thus, in SHE the magicity of a particular
neutron (proton) number very much depends on the number of protons
(neutrons) present in the nucleus.

From the above discussions we conclude that the E-RMF parameter set G2
clearly points out towards the robust double magic character of the
combinations ($N=172,Z=120$) and ($N=258,Z=120$). Similar conclusions
are obtained with the G1 and NL3 parameter sets \cite{sil04}. Also, for
the particular case of the new set G2 the combinations $Z=114$ and
$Z=120$ with $N=184$ show evidences of a shell closure, although less
strong than in the previous cases. A double shell closure in the
($Z=114,N=184$) nucleus has been traditionally predicted by the
macroscopic-microscopic models \cite{mol94}.

\subsection{Shell corrections}

The indicators discussed in the previous section, which allow one to
identify the doubly magic nuclei, correspond to energy differences
between neighbouring nuclei. However, they do not have a direct
connection with the shell corrections which stabilize a given SHE
against fission. In a liquid droplet model picture SHE are unstable
against spontaneous fission due to the fact that the large Coulomb
repulsion cannot be compensated by the nuclear surface tension.
However, SHE may still exist because the quantal shell corrections
generate local minima in the nuclear potential energy surface which
provide additional stabilization. For experimentally known shell
closures (up to $N=126$ and $Z=82$), the shell corrections are
strongly peaked around the magic numbers \cite{kle02}. Thus, the shell
correction energy also represents a test for checking the robustness
of the shell closures.

In Figure 12 we display the total (neutron-plus-proton) shell
correction energies calculated employing the Strutinsky smoothing
procedure \cite{sil04,rin80} for the $Z=114$ and $Z=120$
isotopic chains, using the parameter sets G2 and NL3. The calculations
have been performed in spherical symmetry and thus the computed shell
corrections represent in general an upper bound to the actual ones.
Deformation could bring additional shell stabilization. From the
figure one can see that the isotopic chain of $Z=120$ shows a large
negative shell correction for $N=172$. Another local minimum, less
pronounced, is also found around $N=182-184$. One observes that in
superheavy nuclei the shell correction energy at the shell closures
does not display the very sharp jumps typical of normal mass nuclei.
Rather, the shell corrections depict a landscape of broad areas of
shell stabilized nuclei \cite{kru00,ben2,rei02}. Still, in these areas
the closed shell superheavy nuclei show more negative shell
corrections than their neighbours. Looking at the results for $Z=114$
presented in Figure 12 it can be realized that the shell corrections
in the $Z=114$ isotopes are weaker than for the $Z=120$ chain, which
means less stability. Again, two dips are found at $N=172$ and
$N=184$, although in this case the minimum at $N=184$ is deeper than
the one at $N=172$. The study of the shell correction energy reveals
that the shell stabilized regions of SHE are in good agreement with
the previous predictions for shell closures derived from the analysis
of the energy indicators.

\subsection{Density profiles and spin-orbit potentials}

From a naive point of view one expects that for a large nucleus the
neutron density profile will show a relatively flat region at the
interior, modulated by some wiggles due to shell effects. In the case
of protons one also expects that the Coulomb repulsion will push them
towards the surface so that the proton density will develop a
depression in the center. However, some deviations from this pattern
can arise in superheavy nuclei. To exemplify the situation, we display
in Figure 13 the density profiles of neutrons and protons for the
superheavy nuclides $^{292}120$ (close to the proton drip point),
$^{304}120$, and $^{378}120$ (close to the neutron drip point). For
the purpose of comparison with the situation in smaller nuclei, we
have drawn in the same figure the neutron and proton density profiles
of the isotopes $^{100}$Sn (proton drip line nucleus), $^{132}$Sn, and
$^{176}$Sn (neutron drip line nucleus).

Inspecting the behavior of the tin isotopes with increasing neutron
number $N$, one observes that the neutron density extends
progressively toward the outside, while it shows only a small increase
in the interior. Driven by the proton-neutron interaction, the proton
density also extends more and more to the outside. As a result, since
$Z=50$ is fixed, there is a notable reduction of the average proton
density in the nuclear interior. (See Ref.\ \cite{men99} for a detailed
RHB study of the tin chain.) These general trends are also observed in
the evolution with $N$ of the nucleon densities of the $Z=120$
isotopes displayed on the left side of Figure 13.

It is immediately noted that the neutron density profile of the
$^{292}120$ superheavy nucleus shows an accentuated dip from $r\sim 4$
fm to the origin. As also pointed out in Ref.\ \cite{ben99}, the reason
lies in the fact that for $N=172$ the last filled neutron levels
correspond to a large orbital angular momentum (2g and 1j levels, see
Figure 9). Such orbits are mainly located at the surface and thus
generate the central dip. This is not the case for the system with
$N=184$, where the last occupied neutron levels (4s and 3d) give an
important contribution to the central density. For $N=258$ neutrons
the situation lies somehow in between of the two previous cases due to
the contribution of the 4p and 3f orbitals near the Fermi level. For
the tin isotopes shown in Figure 13, the neutron orbits in the
vicinity of the Fermi energy consist of a bunch of low angular
momentum levels together with the intruder level of higher angular
momentum. The neutron density profiles of the three tin isotopes show
the normal pattern, though for $N=50$ one can recognize a dip around
the center which has the same nature discussed for the superheavy
nucleus $N=172$, now because of the 1g neutron level.

The proton density profiles of the three superheavy isotopes of
$Z=120$ depicted in Figure 13 show a similar pattern. Again, there is
an effect due to the orbitals around the Fermi level. The 2f and 1i
proton orbits of the $Z=120$ nuclides push their
protons away from the center and as a consequence the proton density
develops a pocket around $r\sim 2$ fm. On the other hand, the increase
of the proton density which is observed close to the origin is
reminiscent of the fingerprint of the 3s protons. The proton densities
of the tin isotopes also exhibit a depression near the center. This
time it is occasioned by the 1g proton level, which is the last filled
level for $Z=50$. However, due to the smaller Coulomb repulsion and to
the 2p orbits close in energy to the 1g level, the effect is not as
remarkable as for the proton density of $Z=120$.

We now come to the discussion of the spin-orbit potential in the
superheavy mass region, because it is responsible for the strong
reduction of the spin-orbit splitting of low angular momentum energy
levels found in the self-consistent calculations of
SHE\@. In the relativistic models the spin-orbit force is
automatically included in the interaction from the outset. It appears
explicitly when the lower spinor of the relativistic wave function is
eliminated in favor of the upper spinor. One then obtains a
Schr\"odinger-like equation that contains a spin-orbit term which
reads \cite{fur98a,lal98}:
\begin{eqnarray}
H_{so} & = &
\frac{1}{2 M^2} V_{so}(r) \, {\bf L}\!\cdot\!{\bf S} ,
\label{eqSO2} \\[3mm]
V_{so}(r) & = & \frac{M^2}{{\bar{M}}^2} \frac{1}{r}
\left( \frac{d \Phi}{d r} + \frac{d W}{d r} \right)
+ 2 f_{v} \frac{M}{\bar{M}} \frac{1}{r} \frac{d W}{d r} ,
\label{SOP}
\end{eqnarray}
where we have made $\bar{M} = M - {\textstyle{1\over2}} (\Phi + W)$.
This spin-orbit potential includes the contribution of the tensor
coupling of the $\omega$ meson to the nucleon, which has an important
bearing on the spin-orbit force \cite{fur98,fur98a,est99}. For simplicity,
we have neglected in (\ref{SOP}) the small contributions of the $\rho$
meson and of the Coulomb field which make $V_{so}(r)$ slightly
different for neutrons and protons.

Figure 14 displays the radial dependence of $V_{so}$ given by Eq.\
(\ref{SOP}) for the isotopes of $Z=120$ and $Z=50$, whose density
distributions we have just discussed. A common feature of both the
tin and the superheavy isotopes is that, as expected, the spin-orbit
potential develops a well at the surface region of the nuclear density
distributions. As the system becomes more and more neutron rich, the
depth of this well experiments a gradual reduction and the position of
the bottom of the potential is shifted outwards, which brings about a
weaker spin-orbit force.

There are two distinctive trends of the spin-orbit potential of
superheavy nuclei with respect to normal nuclei. One is that the
spin-orbit well at the surface is less deep in the case of the SHE\@.
The other one is the strong upward bump that $V_{so}$ develops around
$2-4$ fm from the origin. This bump can be explained in connection
with the spatial distribution of the densities. Consider the case of
$N=172$. The nucleon densities are sharply reduced in the center
(Figure 13). To the extent that the meson fields follow the change of
the densities, this causes $V_{so}$ to develop a sharp bump in the
interior of the nucleus \cite{ben99,rei02}. As seen from Figure 14,
the ensuing bump is largest for $N=172$ and decreases with increasing
neutron number as the central region of the neutron densities flattens
out. 

The strong increase of $V_{so}$ near the origin has an outstanding
effect on the energy splitting of spin-orbit partner levels in SHE\@.
Large angular momentum states whose wave function is mostly localized
around the nuclear surface basically feel the attractive part of
$V_{so}$, and thus present normal spin-orbit splittings. However, for
low angular momentum states there is a strong overlap of the wave
function with the positive region of $V_{so}$ which dramatically
reduces the spin-orbit splittings (they may even have the opposite
sign for some parameter sets \cite{ben99}). This effect is corroborated
by looking at the splittings of spin-orbit partners in the SHE spectra
represented in Figure 9.

Little differences are seen in Figure 14 between the predictions of
the G2 and NL3 sets for the spin-orbit potential. The higher effective
mass at saturation of G2 ($M^*_\infty=0.66$) with respect to NL3
($M^*_\infty=0.6$) is compensated by the contribution of the tensor
coupling $f_v$ in the G2 set, since there exists a trade-off between
the size of the $\omega$ tensor coupling and the size of the scalar
field \cite{est99,fur98a}. As noticed in Ref.\ \cite{est01a} the
contribution of $f_v$ to Eq.\ (\ref{SOP}) accounts for roughly
one-third of the spin-orbit strength in the case of the G2 parameter
set. 

Still, inspecting Figure 14, one may observe that the bottom of
$V_{so}$ is located at slightly larger values of $r$ in the G2 set
than in the NL3 set. Also, the spin-orbit well at the surface tends to
be slightly deeper in G2 than in NL3 for the $Z=120$ isotopes (the
opposite happens for tin between $N=50$ and $N=82$). Finally, the
upward bumps appearing in the spin-orbit potential of the SHE at
$r\sim 3$ fm are larger if $V_{so}$ is calculated with NL3. Thus, it
may be expected that the spin-orbit splitting of low angular momentum
energy levels of SHE will be smaller with NL3 than with G2. Since for
the SHE the strength of $V_{so}$ at the surface is a little stronger
and shifted towards slightly larger distances, the spin-orbit
splitting for high angular momentum energy levels should be larger
with G2 than with NL3. These predictions for the splittings of the
energy levels of SHE coming from the small differences in $V_{so}$
obtained with the two considered parameter sets, can be seen in the
neutron single-particle spectra previously displayed.

\section{Concluding remarks}

The effective field theory model derived from the chiral effective
Lagrangian proposed by Furnstahl, Serot and Tang \cite{fur96}
contains, in addition to the standard nonlinear RMF theory, other
nonlinear scalar-vector, vector-vector and tensor couplings consistent
with the underlying symmetries of QCD\@. The Lagrangian is expanded in
powers of the fields and their derivatives and is organized in such a
way that the results are not dominated by the highest-order terms
retained. None of the couplings present in the expansion at a given
order can be eliminated without a symmetry argument. The coupling
constants and some meson masses of the truncated effective Lagrangian,
or equivalently of the energy density functional, are taken as the
effective parameters of the theory and are fitted to measured
observables of a few doubly-magic nuclei (sets G1 and G2). Thus, the
model can be interpreted as a covariant formulation of density
functional theory where the effective functional approximates the
exact (unknown) energy functional of the many-body nuclear system by
expanding it in powers of auxiliary meson fields.

We have checked the extrapolations of the model to the infinite medium
by comparing its results with the Dirac-Brueckner-Hartree-Fock
calculations of nuclear and neutron matter above the normal saturation
density. The sets G1 and G2 derived from the effective field theory
predict a rather soft equation of state both around saturation and at
high densities. These trends are in agreement with the output of the
DBHF theory and recent experimental observations from energetic heavy
ion collisions \cite{stu01,dan02}. At variance with this, the standard
nonlinear RMF parameter sets produce an EOS which is much too stiff
with increasing density because of the restricted density dependence
of the meson fields in the model.

We have analyzed the predictive power of the E-RMF model for
describing finite nuclei in extreme conditions, like those
found at the border of the nuclear chart. The study of
exotic nuclei requires to include the pairing correlations. Our
approach consisted of a modified BCS approximation with constant
pairing matrix elements where the continuum is represented through the
quasi-bound levels retained by their centrifugal barrier. It simulates
the effects of the coupling to the continuum which are incorporated in
more sophisticated treatments of the pairing correlation, like the
relativistic Hartree-Bogoliubov approach or the resonant
continuum-coupling BCS calculations that take into account the width
of the resonant states \cite{dob84,san03}.

We studied very neutron-rich Zr isotopes up to the neutron drip line.
Previous RHB calculations with the NL-SH set predict that these nuclei
exhibit a giant halo \cite{men98}. This effect consists of a noticeable
enhancement of the rms radius of the neutron density distribution due
to the progressive occupancy of tiny bound levels above the
$1h_{11/2}$ level which closes the shell with $N=82$ neutrons. Our
E-RMF calculations with the G2 parameter set produce comparable
results to those obtained with RHB calculations with the NL-SH set. It
is known that for stable medium and heavy nuclei the excitation energy
of the isoscalar giant monopole resonance follows an $A^{-1/3}$ law.
Non-relativistic calculations of the breathing mode in some
neutron-rich nuclei showed that the RPA strength becomes broader as
compared with normal isotopes and that a large amount of strength
concentrates in the low-energy region, because of the transitions from
weakly bound levels to the continuum. We have performed constrained
Hartree calculations with G2 for the Zr chain. The results predict
that the finite nuclei incompressibilities and the average excitation
energy of the breathing mode should fall down steeply for the isotopes
with the largest isospin content. This fact would reflect an
enhancement of the low-energy part of the RPA strength when the nuclei
approach the neutron drip line.

Moving to another edge of the nuclear landscape, we paid attention to
the predictions of the E-RMF sets for the next spherical magic numbers
beyond $Z=82$ and $N=126$. The determination of the shell closures in
superheavy elements is more involved than in normal nuclei where a
large gap in the level spectra usually signals the possible magic
numbers, and one has to study other indicators like two-nucleon
separation energies, two-nucleon shell gaps, and the pairing gap. We
also analyzed the shell correction energy, which provides a
stabilizing effect to superheavy nuclei against fission. Our spherical
prospect must be understood in connection with the occurrence of
double shell closures. Additional deformed shell closures may appear
in the analyzed region of superheavy elements. We contrasted the
density distributions and spin-orbit potentials of superheavy nuclei
against those of normal mass nuclei, with some selected examples from
near the proton drip line to near the neutron drip line. The G2
parameter set predicts strong shell closures for $Z=120$ in
combination with $N=172$ and $N=258$. Again, the E-RMF predictions for
finite systems are in agreement with the results obtained with the
fine tuned standard RMF parameter sets. Interestingly enough, the G2
set presents some indications of a shell closure for $Z=114$ and
$N=184$, which the macroscopic-microscopic models have traditionally
predicted to be a doubly magic nucleus.

In concluding we remark that we tried to show that in the quest for a
unified mean field description of finite nuclei, even in exotic
regions of the nuclear landscape, as well as of nuclear and neutron
matter up to densities several times above the saturation density, the
relativistic mean field model motivated by effective field theory is
one reliable candidate. Surely, this is not the only way of obtaining
such a unified description. Other implementations of effective field
theory for the nuclear many-body problem, formulations of the
relativistic model with inclusion of other mesons, or with
density-dependent coupling vertices, may probably deliver also a
consistent description similarly to the E-RMF model. Nevertheless,
with the present findings we hope to have illustrated the
plausibility and promising potential of the application of effective
field theory methods to the nuclear structure problem.

\acknowledgments
T. S, M. C., and X. V. acknowledge financial support from the DGI
(Ministerio de Ciencia y Tecnolog{\'\i}a, Spain) and FEDER under grant
BFM2002-01868 and from DGR (Catalonia) under grant 2001SGR-00064. 
T. S. also thanks the Spanish Education Ministry grant SB2000-0411.

\pagebreak

%
%
%
\begin{table}
\caption[]{Parameters of the relativistic interactions used in this
work, in dimensionless form, and their saturation properties in
nuclear matter (energy per particle, density, incompressibility,
effective mass, and symmetry energy coefficient).}
\centering
\vspace*{1.cm}
\begin{ruledtabular}
\begin{tabular}{ccccccccc}
 & &  G1 &&  G2 && NL3 \\
\hline
$m_{s}/M$     & & 0.540 & & 0.554 & &  0.541\\
$g_{s}/4\pi$  & & 0.785 & & 0.835 & &  0.813\\
$g_{v}/4\pi$  & & 0.965 & & 1.016 & &  1.024\\
$g_\rho/4\pi$ & & 0.698 & & 0.755 & &  0.712\\
$\kappa_3$    & & 2.207 & & 3.247 & &  1.465\\
$\kappa_4$    & & $-$10.090 & & 0.632 & & $-$5.668\\
$\zeta_0$     & & 3.525 & & 2.642 & &    0.0\\
$\eta_1$      & & 0.071 & & 0.650 & &    0.0\\
$\eta_2$      & & $-$0.962 & & 0.110 & &    0.0\\
$\eta_\rho$   & & $-$0.272 & & 0.390 & &    0.0\\
$\alpha_1$    & & 1.855 & & 1.723 & &    0.0\\
$\alpha_2$    & & 1.788 & & $-$1.580 & &    0.0\\
$f_{v}/4$     & & 0.108 & & 0.173 & &    0.0\\
$f_\rho/4$    & & 1.039 & & 0.962 & &    0.0\\
$\beta_s$     & & 0.028 & & $-$0.093 & &    0.0\\
$\beta_v$     & & $-$0.250 & & $-$0.460 & &    0.0\\
\hline
$a_{v}$ (MeV) & & $-$16.14 & & $-$16.07 & &  $-$16.24\\
$\rho_\infty$ (fm$^{-3}$) & & 0.153 & & 0.153 & &  0.148\\
$K$ (MeV)     & & 215.0 & & 215.0 & & 271.5\\
$M^{*}_{\infty}/M$ & & 0.634 & & 0.664 & &  0.595\\
$J$ (MeV)     & &  38.5 & & 36.4 & &  37.40\\
\end{tabular}
\end{ruledtabular}
\end{table}

%

\begin{figure}
\includegraphics[width=0.9\linewidth]{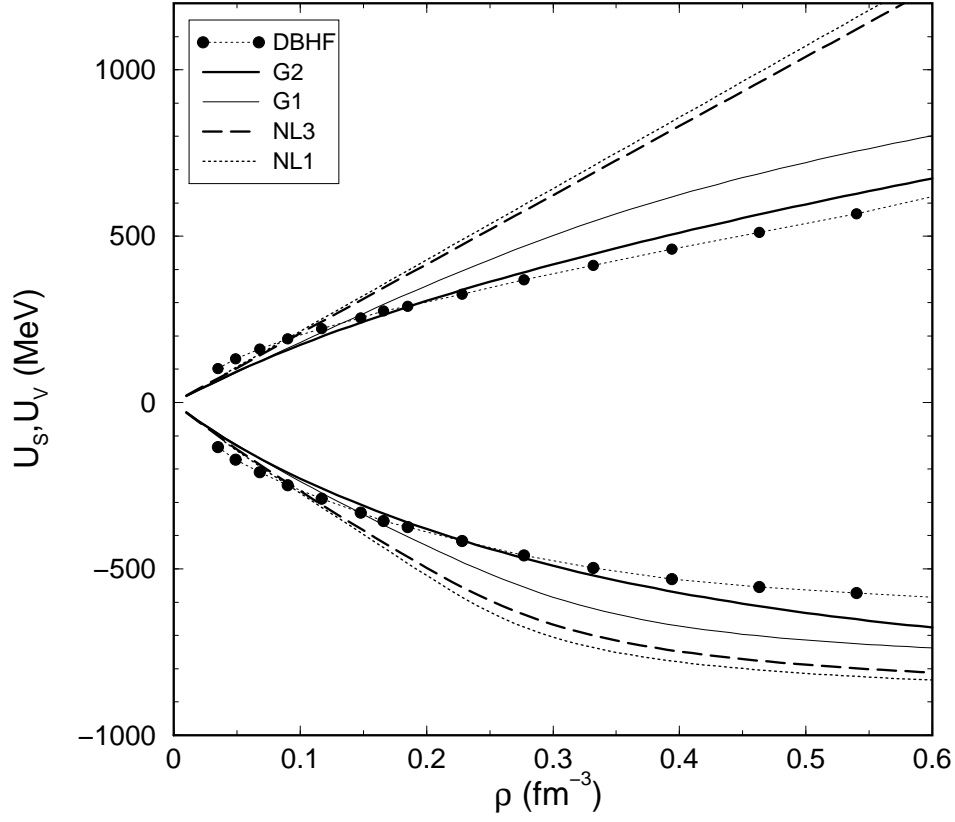}
\\
\caption[]{
The density dependence of the scalar ($U_s$) and vector ($U_v$)
self-energies using the relativistic parameter sets NL1, NL3, G1, and
G2 is compared with the result of the DBHF theory \cite{bro90}.}
\end{figure}

\newpage

\begin{figure}[ht]
\includegraphics[width=0.9\linewidth]{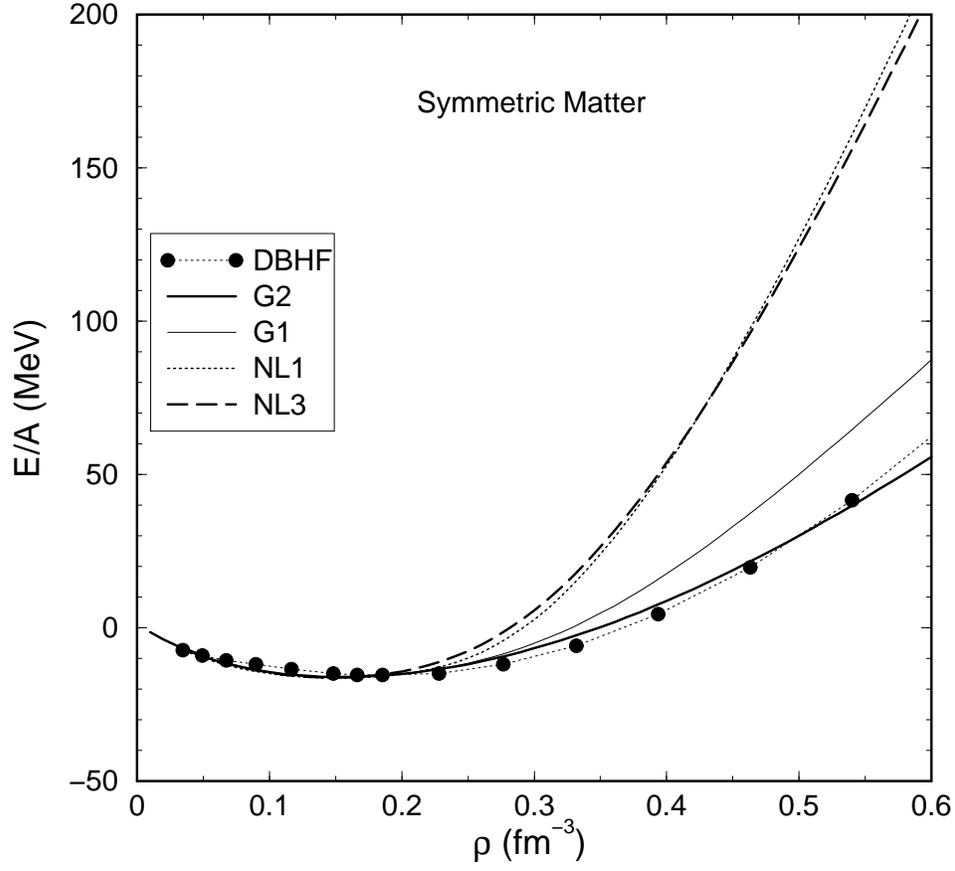}
\\
\caption[]{
Energy per particle of symmetric nuclear matter for the same cases as
in Figure 1.}
\end{figure}

\newpage

\begin{figure}[ht]
\includegraphics[width=0.9\linewidth]{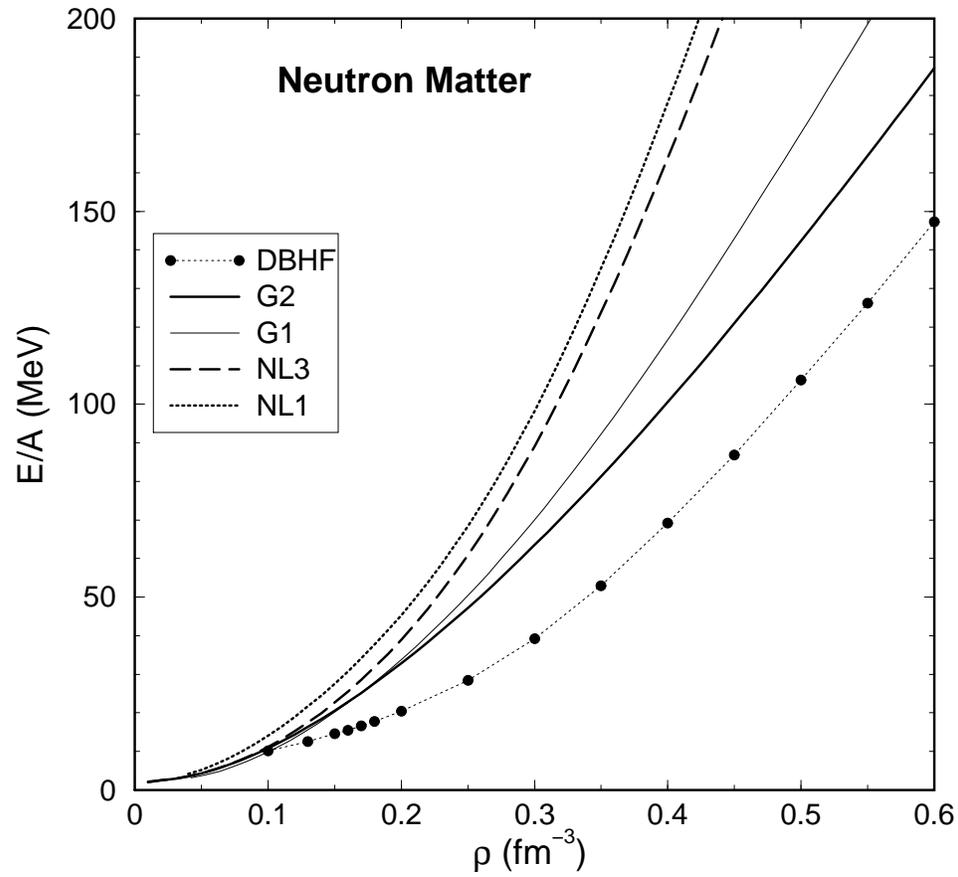}
\\
\caption[]{
Energy per particle of neutron matter for the same cases as in 
Figure 1.}
\end{figure}

\newpage

\begin{figure}[ht]
\includegraphics[width=0.9\linewidth]{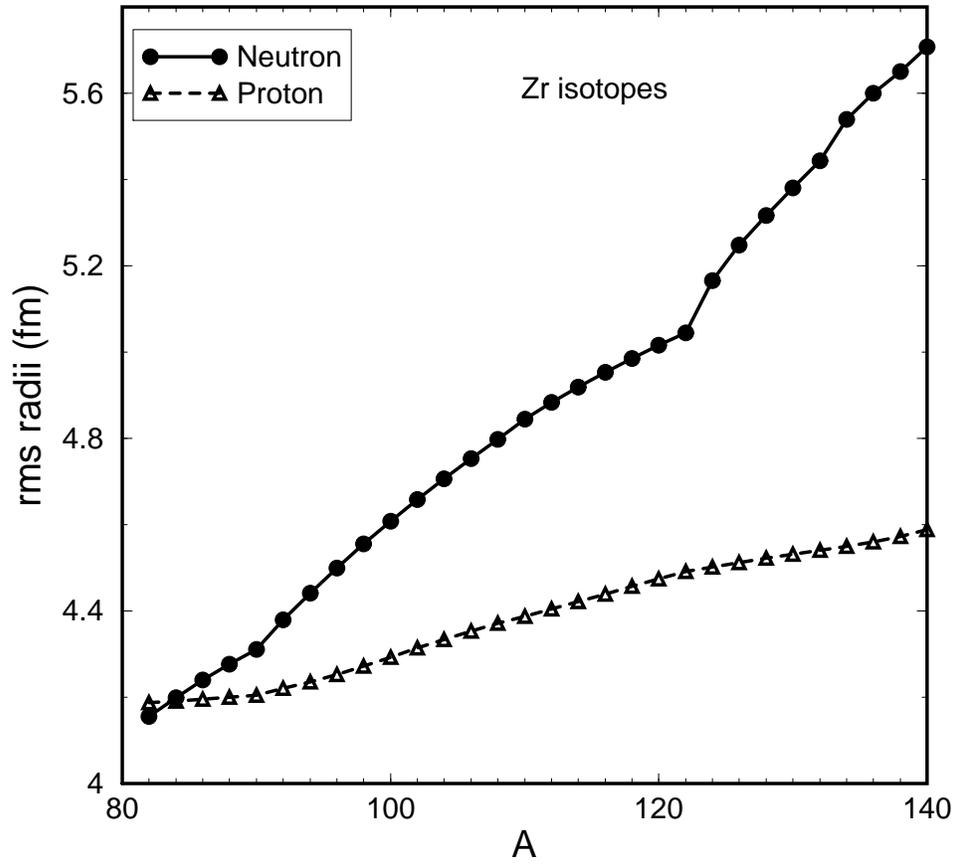}
\\
\caption[]{
Neutron and proton rms radii calculated with the G2 set for the Zr
isotopic chain as a function of the mass number.}
\end{figure}

\newpage

\begin{figure}[ht]
\includegraphics[width=0.9\linewidth]{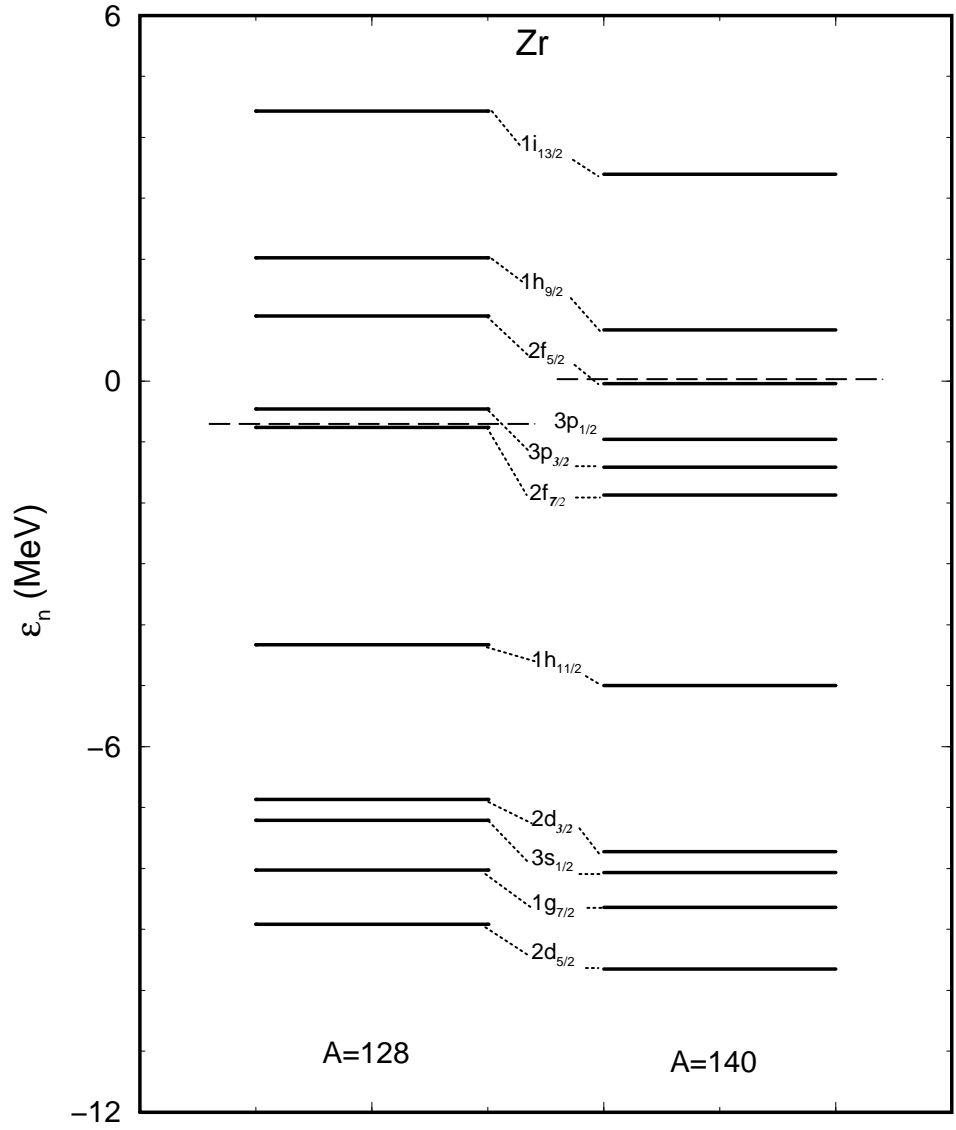}
\\
\caption[]{
Neutron spectra of the $^{128}$Zr and $^{140}$Zr isotopes as predicted
by the G2 set.}
\end{figure}

\newpage

\begin{figure}[ht]
\includegraphics[width=0.8\linewidth,angle=270]{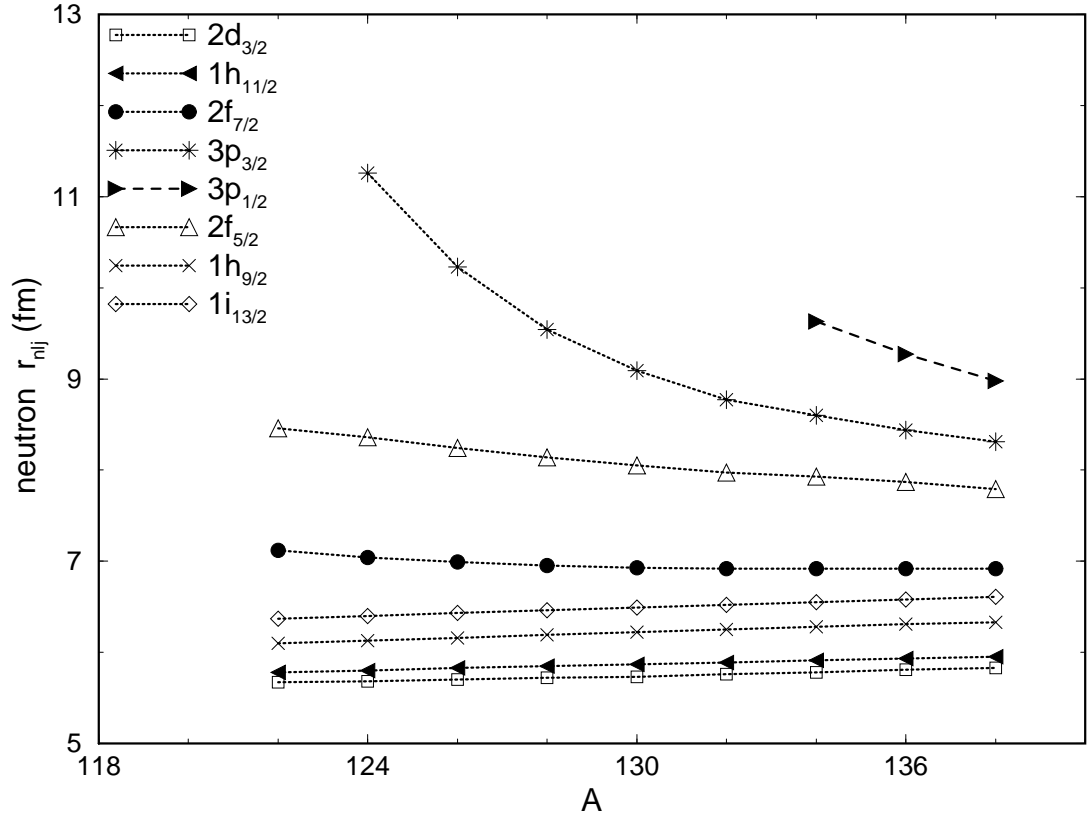}
\\
\caption[]{
Change with mass number of the rms radii of various neutron orbits
for Zr isotopes calculated with the G2 force parameters.}
\end{figure}

\newpage

\begin{figure}[ht]
\includegraphics[width=0.7\linewidth,angle=270]{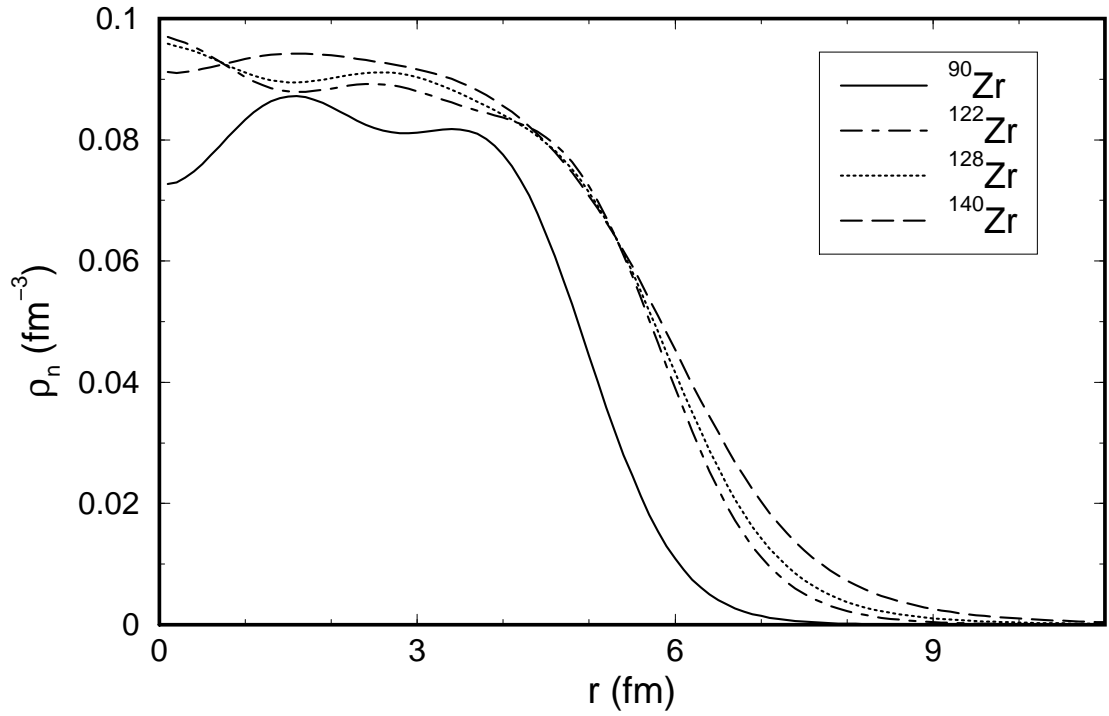}
\\
\caption[]{
Radial dependence of the neutron densities of Zr isotopes.}
\end{figure}

\newpage

\begin{figure}[ht]
\includegraphics[width=0.8\linewidth,angle=270]{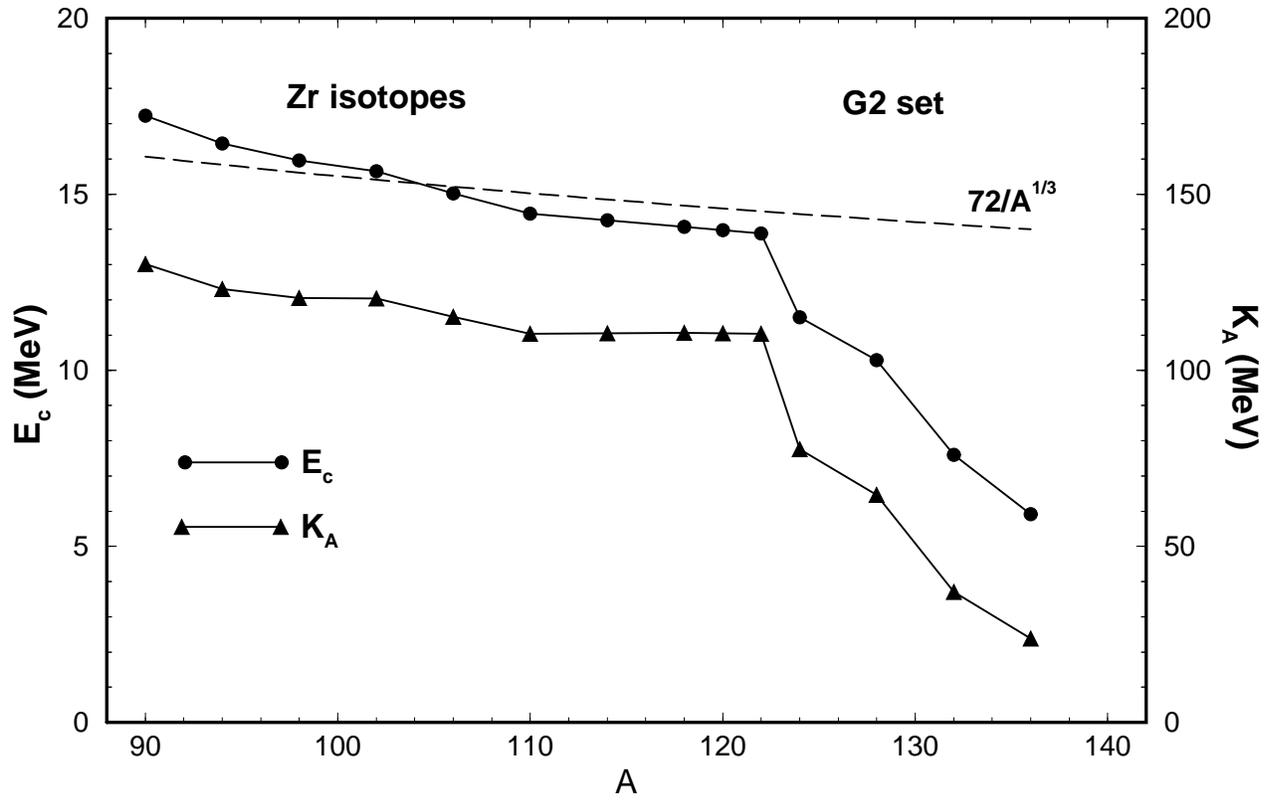}
\\
\caption[]{
Average energy $E_c$ of the isoscalar giant monopole resonance and
finite nucleus incompressibility $K_A$ obtained from constrained
calculations with the G2 set for Zr isotopes.}
\end{figure}

\newpage

\begin{figure}
\includegraphics[width=0.8\linewidth,angle=270]{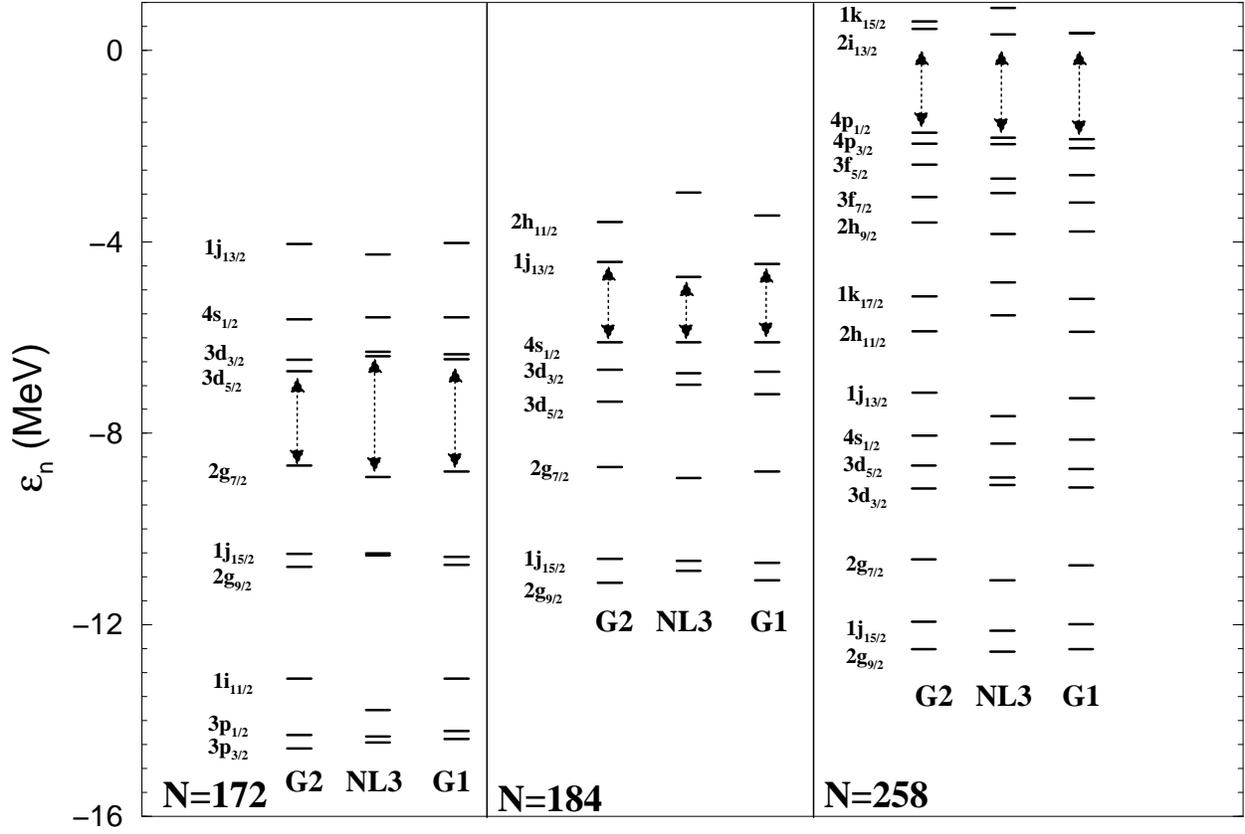}

\mbox{  }  \\

\caption[]{
Single-particle neutron spectra in the vicinity of the Fermi
level for the superheavy isotopes $^{292}120$, $^{304}120$, and
$^{378}120$ computed with the G1, G2 and
NL3 parameter sets.}
\end{figure}

\newpage

\begin{figure}[ht]
\includegraphics[width=0.9\linewidth]{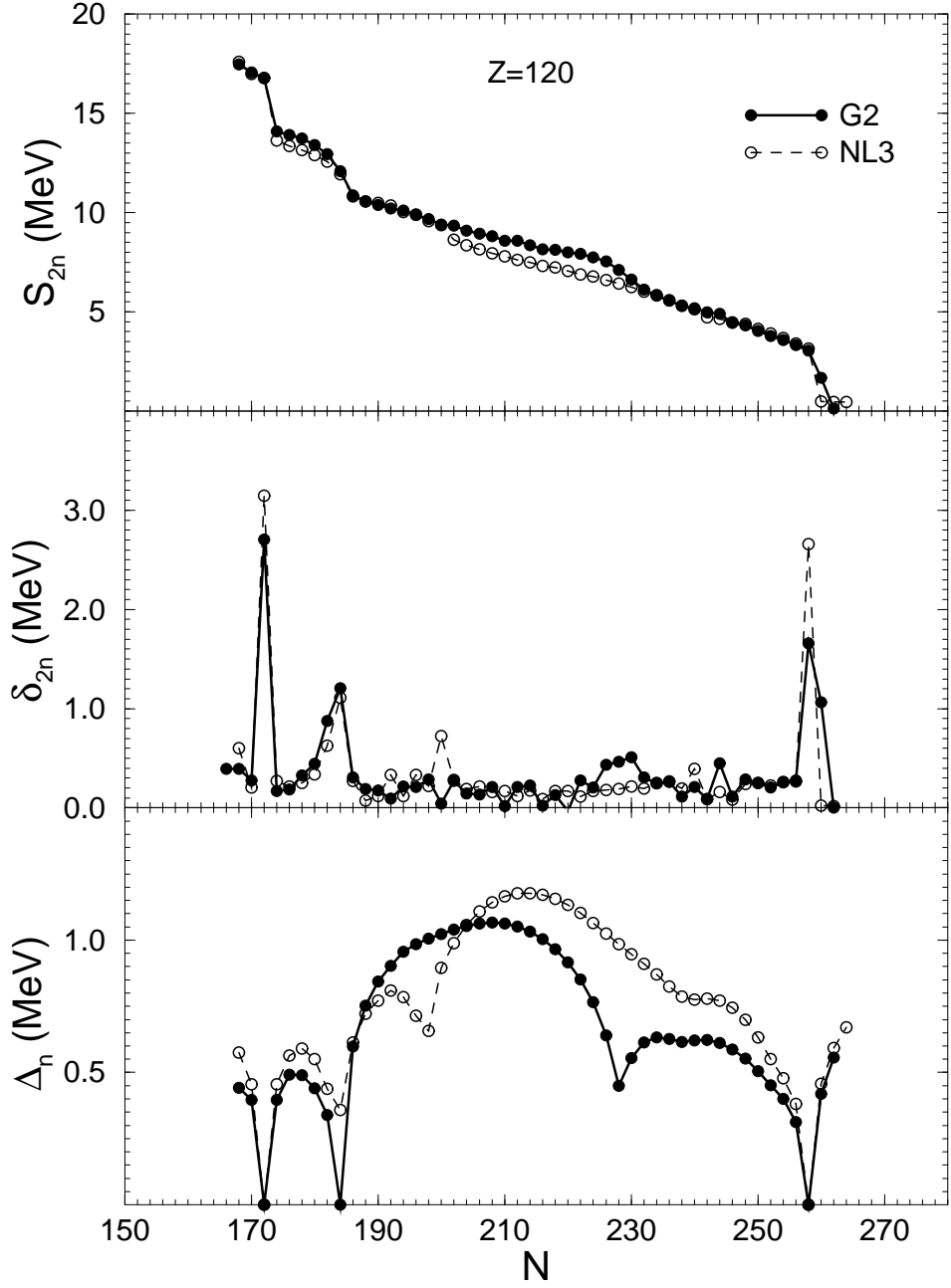}
\\
\caption[]{
Two-neutron separation energy $S_{2n}$, two-neutron shell gap
$\delta_{2n}$, and neutron average pairing gap $\Delta_n$ for $Z=120$
isotopes obtained from spherical calculations with the relativistic
parameter sets G2 and NL3. The proton average pairing gap $\Delta_p$
vanishes in the whole isotopic chain.}
\end{figure}

\newpage

\begin{figure}[ht]
\includegraphics[width=0.9\linewidth,angle=270]{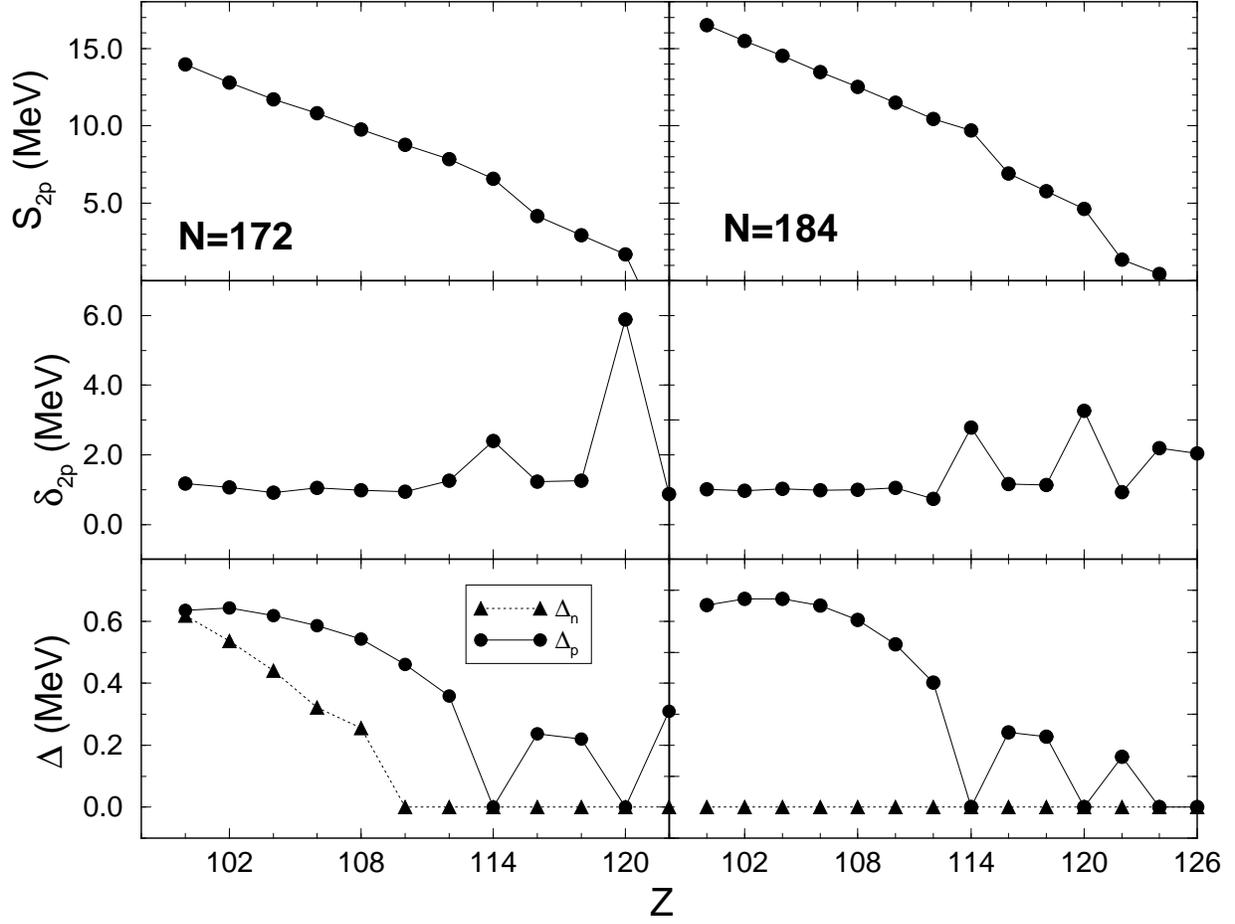}
\\
\caption[]{
Two-proton separation energy $S_{2p}$, two-proton shell gap
$\delta_{2p}$, and proton $\Delta_p$ and neutron $\Delta_n$ average
pairing gaps for $N=172$ and $N=184$ isotones obtained from spherical
calculations with the parameter set G2.}
\end{figure}

\newpage

\begin{figure}[ht]
\includegraphics[width=1.\linewidth]{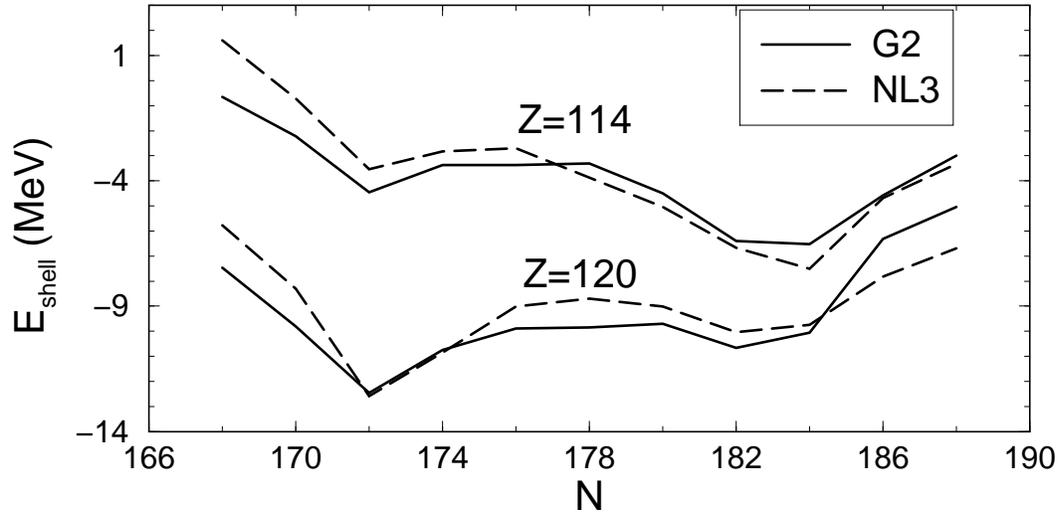}
\\
\caption[]{
Total shell correction energy for $Z=114$ and $Z=120$ isotopes
in spherical calculations with the G2 and NL3 sets.}
\end{figure}

\newpage

\begin{figure}[ht]
\includegraphics[width=0.9\linewidth]{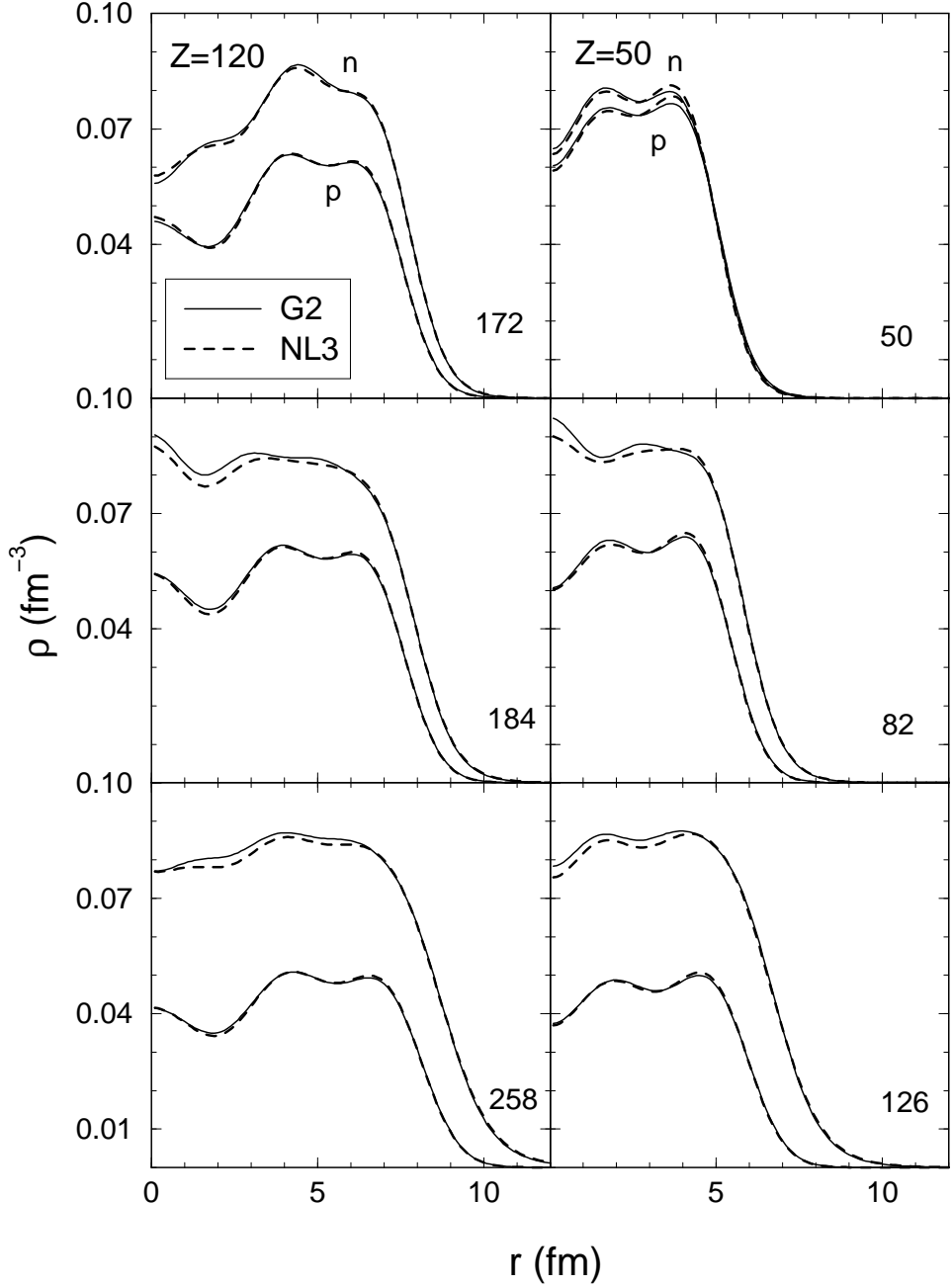}
\\
\caption[]{
The radial neutron and proton density distributions predicted by the
sets G2 and NL3 for the $Z=120$ superheavy isotopes with $N=172$,
184, and 258 (left panels), in comparison with the results for
the tin isotopes with $N=50$, 82, and 126 (right panels).}
\end{figure}

\newpage

\begin{figure}[ht]
\includegraphics[width=0.9\linewidth]{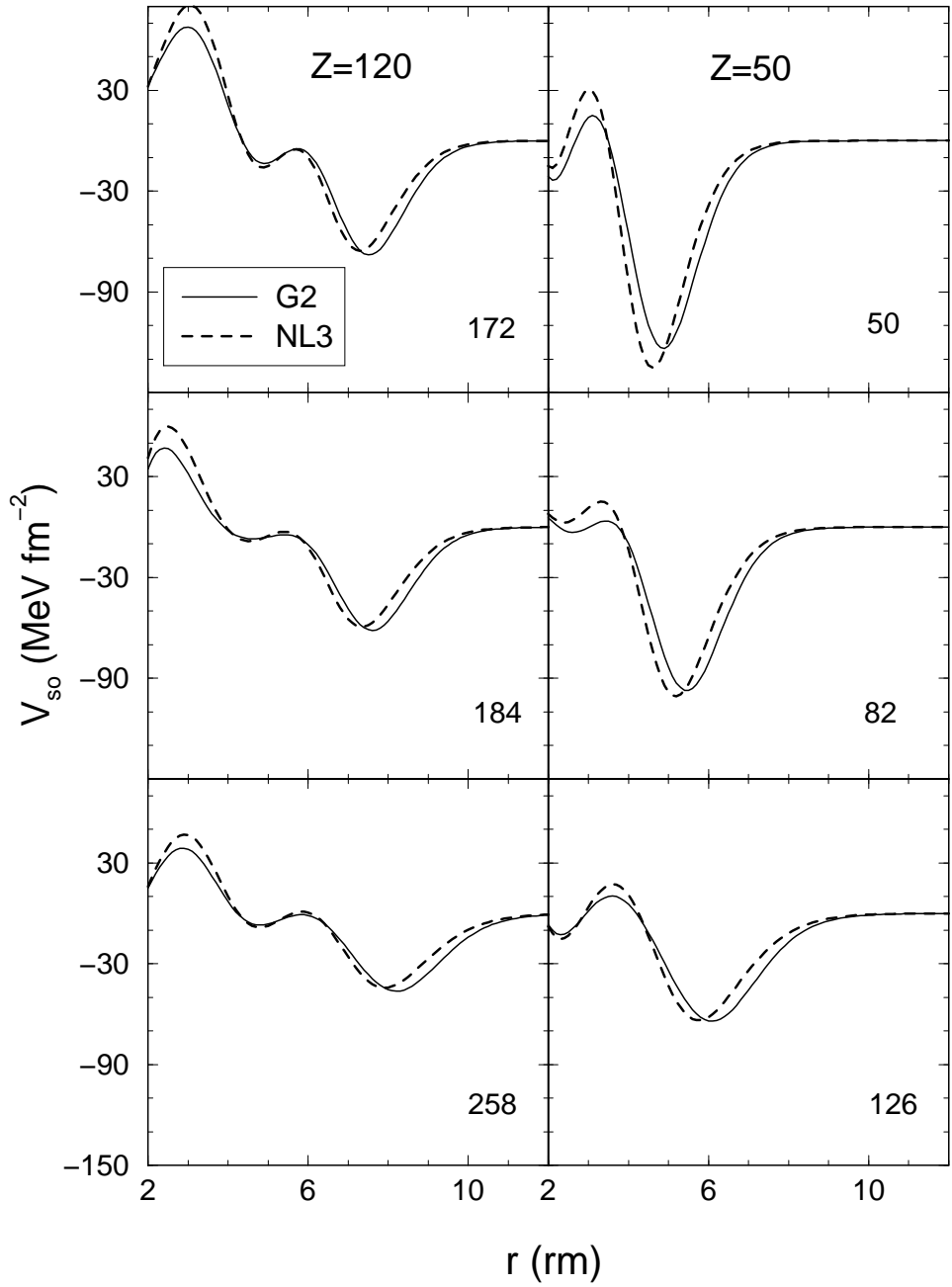}
\\
\caption[]{
Same as Figure 13 but for the radial dependence of the spin-orbit
potential $V_{so}$ defined in Eq.\ (\ref{SOP}).}
\end{figure}

\end{document}